\documentclass[twocolumn,trackchanges]{aastex63}
\usepackage{amsmath}

\providecommand{\e}[1]{\ensuremath{\times 10^{#1}}}
\providecommand{\HII}{H~{\footnotesize II}}	% for H II regions
\providecommand{\OIII}{[O~{\footnotesize III}]}	% for [O III] lines
\providecommand{\OII}{[O~{\footnotesize II}]}	% for [O II] lines
\providecommand{\SII}{[S~{\footnotesize II}]}	% for [S II] lines
\providecommand{\NII}{[N~{\footnotesize II}]}	% for [N II] lines
\providecommand{\NeIII}{[Ne~{\footnotesize III}]}
\providecommand{\HA}{H$\alpha$}			% for H-Alpha
\providecommand{\HB}{H$\beta$}			% for H-Beta
\providecommand{\HG}{H$\gamma$}		% for H-Gamma
\providecommand{\R}{$R_{23}$}			% for R_23
\defcitealias{Paper1}{SFACT1}
\defcitealias{Paper2}{SFACT2}

\shorttitle{The Star Formation Across Cosmic Time (SFACT) Survey.   III.}
\shortauthors{Carr, et al.}

\begin{document}

\title{The Star Formation Across Cosmic Time (SFACT) Survey. III. Spectroscopy of the Initial Catalog of Emission-Line Objects}

\correspondingauthor{David J. Carr}
\email{carrdj@indiana.edu}

\author[0000-0002-4876-5382]{David J. Carr}
\affiliation{Department of Astronomy, Indiana University, 727 East Third Street, Bloomington, IN 47405, USA}

\author[0000-0002-5513-4773]{Jennifer Sieben}
\affiliation{Department of Astronomy, Indiana University, 727 East Third Street, Bloomington, IN 47405, USA}

\author[0000-0001-8483-603X]{John J. Salzer}
\affiliation{Department of Astronomy, Indiana University, 727 East Third Street, Bloomington, IN 47405, USA}

\author[0000-0001-6776-2550]{Samantha W. Brunker}
\affiliation{Department of Astronomy, Indiana University, 727 East Third Street, Bloomington, IN 47405, USA}

\author[0000-0002-7026-1340]{Bryce Cousins}
\affiliation{Department of Astronomy, Indiana University, 727 East Third Street, Bloomington, IN 47405, USA}
\affiliation{Department of Physics, The Pennsylvania State University, University Park, PA 16802, USA}
\affiliation{Institute for Gravitation and the Cosmos, The Pennsylvania State University, University Park, PA 16802, USA}

\begin{abstract}

The Star Formation Across Cosmic Time (SFACT) survey is a new narrow-band survey designed to detect emission-line galaxies (ELGs) and quasi-stellar objects (QSOs) over a wide range of redshifts in discrete redshift windows. The survey utilizes the WIYN 3.5m telescope and the Hydra multi-fiber positioner to perform efficient follow-up spectroscopy on galaxies identified in the imaging part of the survey. Since the objects in the SFACT survey are selected by their strong emission lines, it is possible to obtain useful spectra for even the faintest of our sources (r $\sim$ 25). Here we present the 453 objects that have spectroscopic data from the three SFACT pilot-study fields, 415 of which are confirmed ELGs. The methodology for processing and measuring these data is outlined in this paper and example spectra are displayed for each of the three primary emission lines used to detect objects in the survey (\HA, \OIII$\lambda 5007$, and \OII$\lambda 3727$).   Spectra of additional QSOs and non-primary emission-line detections are also shown as examples. The redshift distribution of the pilot-study sample is examined and the ELGs are placed in different emission-line diagnostic diagrams in order to distinguish the star-forming galaxies from the active galactic nuclei. 

\end{abstract}

\keywords{galaxies: high-redshift --- galaxies: star formation --- galaxies: abundances --- galaxies: evolution --- techniques: spectroscopic --- astronomical databases: surveys}

\section{Introduction} \label{sec:intro}
The Star Formation Across Cosmic Time (SFACT) survey is a new imaging and spectroscopic survey which uses narrow-band (NB) filters to detect large numbers of emission-line sources at a wide range of redshifts. The survey utilizes three primary emission lines to detect objects in its target fields: \HA, \OIII$\lambda 5007$, and \OII$\lambda 3727$. For these primary lines, the survey collects a diverse sample of galaxies that spans the redshift range from the local universe out to z~=~1. Quasi-stellar object (QSO) detections using UV emission lines push that redshift range to z~=~5 and beyond.

The survey's methodology draws from the various emission-line galaxy (ELG) surveys that have come before it. Early ELG surveys utilized objective-prism spectroscopy to select candidates (e.g., \citealt{1975ApJ...202..591S, 1977ApJS...34...95M, 1981ApJS...45..113M, 1982ApJ...258L..11S, 1983ApJS...51..171P, 1983ApJ...272...68W, 1983Afz....19...29M, 1994ApJS...95..387Z, 1996ApJS..105..343Z, 1999A&AS..135..511U, 2000A&AS..142..417H, 2000AJ....120...80S, 2001AJ....121...66S, 2002AJ....123.1292S}) and more recent surveys utilize narrow-band imaging data (e.g., \citealt{1993ApJ...412..524B, 2004AJ....127.1431R, 2007ApJ...668..853K, 2010AJ....139..279W, 2011ApJ...726..109L, 2012AJ....143..145K, 2012MNRAS.420.1926S, 2013MNRAS.428.1128S, 2015MNRAS.453..242S, 2019ApJ...880....7C, 2020AJ....160..242S, 2020MNRAS.493.3966K, 2021ApJS..253...39W}). SFACT expands and complements these existing surveys by using three custom narrow-band filters that enable the detection of ELGs out to cosmologically interesting redshifts. The goal of the survey is to discover a statistically complete sample of ELGs that is useful for a broad range of science applications (see \citealt{Paper1}, hereafter \citetalias{Paper1}).

SFACT is being introduced in a series of three papers that focus on the initial release of the survey data from three pilot-study fields.  The first of the three introductory papers is \citetalias{Paper1}, which presents an overview of SFACT's goals, motivations, and the planned scope of the overall survey.  It also discusses the different types of ELGs that SFACT is designed to discover.  SFACT1 presents early results from the survey, describing the properties of the sample of ELGs detected in the pilot-study fields.  In addition, \citetalias{Paper1} details some of the planned science applications the survey data can be used for.  Imaging and spectroscopic data for several newly discovered objects are presented to illustrate the nature of the survey constituents. 

The second SFACT paper is \citet{Paper2} (hereafter \citetalias{Paper2}). It presents the imaging portion of the survey, discussing the methodology for acquiring and processing the imaging data and the details of target selection. It presents catalogs of ELGs detected in the pilot-study fields, as well as images for a set of example objects.  It analyzes the photometric properties of the pilot-study sample as well as how the survey's selection parameters relate to narrow-band flux.

In the current paper we present the spectroscopic portion of the survey for the pilot-study fields.  SFACT was conceived from the start as a dual NB imaging plus spectroscopic survey (see \citetalias{Paper1}).  While the detection of the ELG candidates comes entirely from the NB imaging portion of the survey, the spectroscopic follow-up provides the information necessary for carrying out many of the proposed science applications described in \citetalias{Paper1}.  Fundamentally, the spectra allow us to confirm the ELG nature of our sources, identify the emission line that the survey has detected, and provide an accurate redshift for determining distant-dependent quantities such as luminosities and star-formation rates.  The nebular spectra also allow us to determine accurate absorption corrections for each galaxy based on their observed Balmer decrements.  For galaxies detected via the \HA\ and \OIII$\lambda 5007$ lines, we are also able to measure metal abundances for many sources.  Finally, the spectra obtained for our pilot-study candidates have been extremely valuable for evaluating the survey selection method, allowing us to  modify our procedures to improve the accuracy and efficiency of the survey.

The contents of SFACT3 are presented as follows.  First, we provide a brief overview of the imaging portion of the survey, in order to help place the rest of this paper into context.
In Section \ref{sec:observations}, the instrumentation and procedures used to complete the spectroscopic observations are illustrated. The details of the spectral reduction and line measurement software are explained in Section \ref{sec:spectra_processing}. Finally, Section \ref{sec:results} presents the tabulated spectroscopic pilot-study data as well as example spectra from the survey to illustrate the variety of objects detected. The properties of the ELG sample are displayed in a redshift histogram and emission-line diagnostic diagrams. 

For all of the SFACT papers, a standard $\Lambda$CDM cosmology with $\Omega_m$~=~0.27, $\Omega_\Lambda~=~0.73$, and H$_0~=~70$ kms$^{-1}$ Mpc$^{-1}$ is assumed.

\section{Overview of SFACT Imaging Survey} \label{sec:overview}

The narrowband imaging portion of SFACT is summarized in SFACT1 and described in great detail in SFACT2.  Here we provide an overview of the observations and object selection method, to help frame the contents of the current paper. 

All imaging observations for SFACT are obtained using the One-Degree Imager  (ODI; \citealt{2010SPIE.7735E..0GH}) camera on the WIYN 3.5m telescope\footnote{The WIYN Observatory is a joint facility of the University of Wisconsin–Madison, Indiana University, NSF’s NOIRLab, the Pennsylvania State University, Purdue University, and the University of California, Irvine.}.  ODI was a field-of-view of 40 by 48 arcmin, with a native pixel scale of 0.11 arcsec pixel$^{-1}$.  Observations are obtained for each survey field through three broadband filters (SDSS {\it gri}) and three narrowband filters.  In order to eliminate the many chip gaps present in the ODI focal plane, all observations are acquired using a 9-point dither pattern.   The exposure times used for each image in the dither sequence is 120 s for the broadband images and 600 s for the narrowband images.  For more information concerning the data acquisition and imaging processing, see \citetalias{Paper2}. 

 The three SFACT narrow-band filters used to detect potential ELGs are custom filters designed and fabricated for the survey. The filters all have a bandwidth of $\sim$90 \AA, and have central wavelengths of 6950 \AA\ (referred to as NB1 throughout the remainder of this paper), 6590 \AA\  (NB2), and 7460~\AA\ (NB3).   The redshift ranges of the detected ELGs depends on both which filter the signal is seen in and the specific emission line present in the filter (e.g., \HA, \OIII$\lambda 5007$, \OII$\lambda 3727$, etc.): SFACT1 tabulates these ranges for the primary emission lines detected in our survey.  Future expansion of the survey will add additional NB filters at $\sim$8120 \AA\ and $\sim$9120 \AA.  
 
 The ELG candidates are selected from the images by comparing the fluxes measured in each NB filter with a corresponding measurement made in suitably scaled broadband images (the sum of the {\it r} and {\it i} filters for NB1 and NB2, and {\it i} for NB3).  Objects are selected as SFACT candidates if they possess an excess of flux in the NB image amounting to a magnitude difference $\Delta$m = 0.4 mag, as long as the flux excess is statistically significant (greater than 5$\sigma$).
 
 The survey is quite deep, with a median g-band value for candidate ELGs of 23.2.  The limiting emission-line flux level of the resulting ELG catalog is $\sim$1.0 $\times$ 10$^{-16}$ erg s$^{-1}$ cm$^{-2}$.
    
\section{SFACT Spectroscopic Observations} \label{sec:observations}
%Telescope discussion and Imaging discussion
All spectroscopic data obtained for SFACT were acquired using the WIYN 3.5m telescope located at Kitt Peak National Observatory. This section will cover the instrumentation and observational procedures associated with the spectroscopic part of the survey.  The spectroscopic data for the SFACT pilot-study fields were obtained during observing runs in November 2017, October 2018, August 2019, October 2019, and October 2021.

\subsection{Instrumentation} \label{subsec:instrumentation}
    %Hydra specifications/justification
    The spectroscopic data for SFACT are taken using Hydra and the Bench Spectrograph. Hydra is a multi-fiber positioner with a field of view of 1.0 degree in diameter. SFACT was developed with the intention of using Hydra on the WIYN 3.5m telescope to carry out follow-up spectroscopy. Hydra is able to place fibers on $60-65$ targets per configuration, which allows SFACT to efficiently gather follow-up spectra for all of its potential targets. Its field of view is a good match to the footprint of the ODI camera.  This makes it an excellent tool for acquiring follow-up spectra for the survey.
    
    %fiber specifications and description of how telescope works
    We use Hydra's red cables for our observations.  Each red fiber subtends 2 arcseconds on the sky meaning that most of the light from our higher redshift sources should be within the diameter of the fiber.  %A mechanical gripper moves each fiber to the desired location on the focal plate and the magnet at the bottom of each fiber button holds it in place.  %Bench Spectrograph section
 Light from each source flows down the fibers and is collected by the Bench Spectrograph which is isolated in a separate room in the lower level of the WIYN facility. For the SFACT survey, we chose to use the 600 @ 10.1 bench grating because it has the highest efficiency across our desired wavelength range. It has 600 grooves per mm, a spectral resolution of 3.35~\AA\ with the red cables, and a dispersion of 1.41~\AA/pixel after binning. Pixels are binned 2 x 2 during readout to increase signal-to-noise without losing any resolution.
    
    %Wavelength range: 
With our chosen spectroscopic setup the observed wavelength range for our follow-up spectra is roughly $4760-7580$~\AA.   A primary criterion for our required wavelength coverage was that it includes the spectral ranges covered by our three NB filters.   This guarantees that the emission line seen in our NB images will be present in our follow-up spectra.

\subsection{Observational Procedure} \label{subsec:observational_procedure}

    %Pointing file logistics
   The set-up of the Hydra fiber positioner is controled by the use of ``pointing files", which contain the coordinates and instructions used by the mechanical gripper to configure each field.   Every SFACT field presented in this paper was observed on multiple nights, each time with different pointing files. Each pointing generally contained between $20-60$ SFACT targets, $12-20$ sky fibers, and $3-7$ Field Orientation Probes (FOPs). FOPs are used to accurately align the telescope to each field and are also used as guide stars. %A minimum of three must be assigned for the telescope to track the field during the observation and keep it locked on and aligned toward the targets. 
 All FOPs stars were selected to have \textit{g}-band magnitudes between 10.5 and 14.0 in the Sloan Digital Sky Survey (SDSS; \citealt{2000AJ....120.1579Y, 2019ApJS..240...23A}), with preference given to stars between 12 and 13 magnitudes. 
    
    %Pointing observations
    Each pointing was generally observed for three, 30 minute exposures. Multiple exposures are taken so that cosmic rays and other artifacts are removed from the data when the images are combined. On nights of reduced transparency, pointings would be observed for four or five different exposures in order to achieve as much depth as possible in the final combined images. Due to the faintness of the objects in the SFACT catalog, observations were rarely carried out while the moon was above the horizon.
    
    %Calibration observations
    A series of bias, dome flat, comparison lamp, and dark-current images were acquired during each night of observing.  Standard stars were generally observed in the beginning, middle, and end of each night of observing and are used to create a nightly sensitivity function as discussed in Section \ref{sec:spectrareduc}.
    
In ideal circumstances, the earliest spectra can be gathered for our fields is one year after the full set of imaging data are obtained.  However, it generally takes longer than one year to get \emph{complete} spectroscopic follow-up of every ELG in these fields. This has been the case for a variety of factors including, but not limited to: the loss of telescope time due to the COVID-19 pandemic, bad weather rendering scheduled observing time useless, and fiber placement limitations in crowded regions of the field. Hence, not all SFACT candidates in the final catalog lists presented in \citetalias{Paper2} have been observed spectroscopically. Despite the lack of completeness  there are enough spectra to get a clear picture of the nature of the SFACT survey and of the various objects discovered through our efforts.

    %Image of finallized spectra
    \begin{figure*}[t]
    \centering
    \includegraphics[width=0.975\textwidth,keepaspectratio]{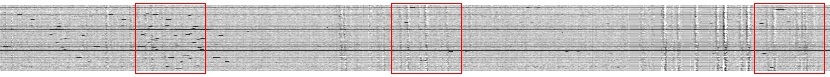}
    \includegraphics[width=0.32\textwidth,keepaspectratio]{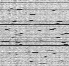}
    \includegraphics[width=0.32\textwidth,keepaspectratio]{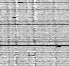}
    \includegraphics[width=0.3246\textwidth,keepaspectratio]{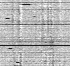}
    \caption{Top: The redward half of the final processed multi-object spectral image from a pointing of SFF15. Each row in the image is a one-dimensional spectrum of an individual SFACT target. The red boxes show the wavelength coverage of the three narrow-band filters. The leftmost box is NB2, the middle box shows NB1, and the right box shows NB3. In this pointing many NB2 candidates were observed. Bottom: The spectral regions covered by SFACT's filters are enlarged and presented in the same order from left to right as mentioned above. \label{fig:spectra}}
    \end{figure*}

\section{Spectral Processing} \label{sec:spectra_processing}

This section will discuss the details of how the spectroscopic data are processed and how we measure the lines in the spectra of our newly discovered ELG sample.

\subsection{Spectral Reductions} \label{sec:spectrareduc}

    %IRAF, part 1, and imcombine
    The first phase in the processing is carried out using the Image Reduction and Analysis Facility ({\tt IRAF}). First, the overscan level is measured and subtracted from each image. Then the biases, darks, and flats are averaged, and biases and darks are subtracted from the data. Next, the three or more science images for each field are median combined to remove cosmic rays and artifacts. The images are median scaled before being combined in order to place the continuum on a similar level between the images. This scaling is based on a skyline-clean region in a brighter object's spectrum.    
    %Hydra package
    The second phase of the processing uses {\tt IRAF}'s {\tt HYDRA} package and {\tt DOHYDRA} task \citep{dohydra}. {\tt DOHYDRA} carries out the following tasks on each multi-fiber spectral image: (1) identifies each spectrum and performs a spectral trace for each fiber; (2) corrects for scattered light; (3) extracts the flat field spectra, using the same extraction aperture as used on each source, and applies them to the science spectra; (4) extracts the comparison lamp spectra and derives a wavelength solution for each fiber; (5) creates a composite sky spectrum and subtracts it from each science spectrum.
    
    %Standard 
    Flux calibrations are performed using the standard stars observed as part of our program.  A sensitivity function is generated for each night and applied to the galaxy spectra.  While accurate spectrophotometry using fibers is notoriously unreliable, our final calibrated spectra should have accurate {\it relative} fluxes, meaning that we can extract reliable emission-line ratios.  Telluric absorption corrections are applied using a well exposed standard star spectrum as a template.  Finally, regions around strong night sky lines are masked to prevent sky residuals from dominating the faint emission lines in our galaxy spectra.  These masked sky lines include [O~{\footnotesize I}]$\lambda$5577, $\lambda$6300, and $\lambda$6363, and NaD among other lines. 
    
%Figure discussion
Figure \ref{fig:spectra} displays portions one of the final spectral images from a single pointing of field SFF15, processed as described above.  The top panel shows the red half of the full image, where only SFACT objects are displayed  (fibers dedicated to extra targets or sky measurements have been removed). Wavelength in the figure increases from left to right, and the spectral range shown covers rest wavelengths of $\sim$6350--7530 \AA.  Each row shows the processed spectrum of a single SFACT galaxy.   The short, dark horizontal lines represent the emission lines in the spectra. Features repeated across multiple rows (vertical) indicate residual flux from night sky lines. The red boxes in the top portion of the figure denote the wavelength ranges of the three SFACT narrow-band filters. NB2 is centered at 6590~\AA, NB1 is at 6950~\AA, and NB3 is at 7460~\AA. 
    
    The bottom panels of Figure \ref{fig:spectra} zoom in on the spectral regions covered by the three filters to display the emission-line detected galaxies in this pointing in more detail.  The majority of the targets in this pointing happened to be detected in the NB2 filter (leftmost lower image), while NB1 (center) and NB3 (right) contain several detections each.   Close inspection of the NB2 sub-image reveals a number of spectra with two emission lines present in the same row.   These are cases where both [\ion{O}{3}] $\lambda$5007 and $\lambda$4959 are included within the filter.

\subsection{Automatic line measurement with {\tt WRALF}} \label{sec:wralf}
    %WRALF
    Identification and measurement  of the emission lines in the spectra is carried out using {\tt WRALF} (WRapped Automated Line Fitting; \citealt{Cousins}). {\tt WRALF} is a python wrapper for a customized version of {\tt ALFA} (Automated Line Fitting Algorithm; \citealt{2016MNRAS.456.3774W}) and operates on the type of multi-spec format files illustrated in Figure \ref{fig:spectra}.  {\tt WRALF} displays each spectrum from the image in turn, and the user is responsible for identifying and marking a single strong emission line.  The code then calls {\tt ALFA}, passing a redshift estimate based on the identified line. {\tt ALFA} fits the continuum of the spectrum by assigning the central pixel in a moving 100 pixel window to be the 25th percentile of flux values within the window.   The code then uses a lookup table of possible emission lines, and attempts to identify and measure all lines that are present in the wavelength range of the spectrum, fitting a Gaussian at the locations of potential lines. 
    
    {\tt ALFA} estimates the uncertainty by subtracting the best-fitting solution from the continuum-subtracted observed spectrum. These residuals are used to calculate the signal-to-noise of each line. If the ratio of signal-to-noise is less than three, {\tt ALFA} does not consider that line to be a real measurement.
    
    The output from {\tt ALFA} is shown to the user who determines if the identified lines are real and, if they are, the spectrum is saved and the identified lines and their properties are recorded. Measured lines with a signal-to-noise greater than or equal to five are used to derive a series of redshift measurements and these measurements are averaged to give the final redshift for the object. The standard deviation of the individual redshift estimates serves as the redshift uncertainty.  Equivalent width (EW) is calculated for each line by taking the flux returned by {\tt ALFA} and dividing by the continuum measured over a small range determined by the full width at half maximum (FWHM). 
    
    All spectral information is merged back into the SFACT table databases for each field. At the end of the process, {\tt WRALF} has measured the redshift of each source and the observed wavelength, flux, error in the flux, EW, and FWHM of each emission line in the spectra that it detects.
    
    While {\tt ALFA} is able to measure narrow emission lines accurately, it struggles to automatically identify lines in two cases. First, {\tt ALFA} has trouble identifying broad emission lines in objects like quasars and Seyfert 1 AGN. These lines are broad enough that {\tt ALFA} mistakes them to be a part of the continuum and the software often does not identify them as emission lines. Second, {\tt ALFA} misses lines in spectra that are of lesser quality or that contain weak signal-to-noise emission lines. 
    
    To remedy this, some objects had to be re-examined using an auxiliary code and their information had to be entered into the data tables separately. The code operates on the two cases differently. In the first case, broad-line objects are flagged during the data processing. These objects are displayed for the user who then measures the missing emission lines manually.
    
    In the second case, objects with weak signal-to-noise emission lines are identified by searching through the data tables and displaying the spectra of objects missing their \NII, \HB, or \OII\ lines. These three emission lines are specifically targeted so that as many sources as possible can be included in emission-line diagnostic diagrams to separate them based on their activity class (see Section \ref{sec:emissionlinediagnosticdiagrams}). Stronger lines like \HA\ and \OIII$\lambda 5007$\ are rarely missed by WRALF. The user then makes a decision for each spectrum if any of these lines are present. 
    
    If a line is found in the expected location then it is categorized in one of two ways. In many cases, the re-examined line is clearly present at the right wavelength but its signal-to-noise is slightly below WRALF's retention limit. These emission lines are measured and the re-examination measurement is labeled as a Category 1 measurement. If a feature is present at the expected wavelength but the feature is weaker in nature and similar in strength to the surrounding noise, it is measured and then flagged as a Category 2 re-examination measurement. In some cases, Category 2 measurements are essentially an upper limit on the line measurement. In other cases, the measurement of the Category 2 lines are simply lines with low signal-to-noise that have measurements that are more dubious than the Category 1 measurements.

\subsection{Limitations of the SFACT Spectra} \label{sec:limitations}

   We point out two important limitations of our follow-up spectra.  The first is the well-known issue that the measurement of accurate absolute fluxes is notoriously difficult when using a fiber-fed instrument like Hydra.  This is due to a number of reasons, which include (1) the fact that each fiber will have a slightly different throughput, (2) observations of flux calibration standard stars are typically only done through one fiber rather than all fibers, and (3) the positioning of the fibers on the focal plain of the telescope is not perfect.  The latter issue can result in imprecise alignment between the fiber and the astronomical source, resulting is the loss of measured flux through the fiber.

   For the current project the situation is exacerbated by the faint nature of the sources and their small angular extent on the sky (often point-like in nature).   A comparison of the emission-line fluxes measured with Hydra with those derived from our NB photometry shows a large scatter, even when the comparison is limited to unresolved objects.   Hence, we do not trust the line fluxes measured in our follow-up spectra.   Our relative fluxes (i.e., emission-line ratios) should be robust, however.

   A second problem is associated with our measured EWs for objects with faint underlying continua.  As detailed in SFACT1 and SFACT2, many of our sources have r-band magnitudes fainter than 24.  The observed variation between the continuum fluxes in the individual sky spectra used to create the composite sky spectrum employed in our sky subtraction has a typical value of $\pm$5-8\%, probably due to fiber-to-fiber throughput variations.   This translates into a characteristic uncertainty in our sky subtraction of around 5 $\times$ 10$^{-18}$ erg s$^{-1}$ cm$^{-2}$ \AA$^{-1}$ in the continuum.   For some of our sources (5-10\%) this sky-subtraction uncertainty results in negative continua and negative EWs.

   The problem is more insidious than the presence of negative EWs.  For many of our faint ELGs, the measured continuum levels after sky subtraction will be positive but have very small values (less than 1 $\times$ 10$^{-18}$ erg s$^{-1}$ cm$^{-2}$ \AA$^{-1}$).   An object with a measured continuum of 5 $\times$ 10$^{-19}$ erg s$^{-1}$ cm$^{-2}$ \AA$^{-1}$ will have an uncertainty in the continuum of a factor of 10, and an uncertainty in the measured EW of a line of the same size.   Hence, even for sources with positive EWs the measured values can be totally unreliable.  One object located in the pilot-study fields has a measured EW for [\ion{O}{3}]$\lambda$5007 of 6800 \AA, which is unphysicaly large.

   The problems with our measured EWs are, unfortunately, inherent in the process of trying to observe very faint objects with a multi-fiber spectrograph.  We are exploring methods to help mitigate the impact on our data.   We stress that this negative continuum issue does not impact our line flux measurements (and hence our line ratios), but it does render at least some of our EWs unreliable.

% \movetabledown= 30mm
%\begin{rotatetable}
    \begin{deluxetable*}{ccccccccccccc}
    \tabletypesize{\scriptsize}
   \tablecaption{SFF01 Spectral Data Table \label{tab:specdata_sfactf01}}
    \tablehead{
    \\
     SFACT Object ID & SFACT Coordinate ID & $\alpha$(J2000) & $\delta$(J2000) & Type & z & Line & log([\ion{N}{2}]/H$\alpha$) & log([\ion{O}{3}]/H$\beta$) & log([\ion{O}{2}]/H$\beta$) \\
    & & degrees & degrees & & & &  \\  %$\frac{\text{erg}}{\text{s } \text{cm}^2}$
    (1) & (2) & (3) & (4) & (5) & (6) & (7) & (8) & (9) & (10) 
    }
    \startdata
 SFF01-NB3-D20110 & SFACT J214123.25+200510.7 & 325.34689 &  20.08630 & SFG &   0.13692 &  6563 & -0.541 $\pm$  0.049 & -0.513 $\pm$  0.123 & \nodata \\
 SFF01-NB3-D20084 & SFACT J214123.34+200509.5 & 325.34723 &  20.08598 & HII &   0.13662 &  6563 & -0.640 $\pm$  0.023 & -0.255 $\pm$  0.017 & \nodata \\
 SFF01-NB3-D19969 & SFACT J214123.61+201118.8 & 325.34839 &  20.18856 & SFG &   0.13678 &  6563 & -0.794 $\pm$  0.101 &  0.466 $\pm$  0.109 & \nodata \\
 SFF01-NB3-B20552 & SFACT J214126.31+195845.4 & 325.35962 &  19.97928 & SFG &   0.14168 &  6563 & -0.462 $\pm$  0.074 & \nodata & \nodata \\
 SFF01-NB2-D19115 & SFACT J214126.46+201342.6 & 325.36023 &  20.22851 & SFG &   0.31981 &  5007 & \nodata &  (0.774 $\pm$  0.147) &  (0.154 $\pm$  0.172) \\
 SFF01-NB1-B20542 & SFACT J214126.46+194344.1 & 325.36023 &  19.72892 & SFG &   0.38529 &  5007 & \nodata &  0.462 $\pm$  0.091 &  0.463 $\pm$  0.106 \\
 SFF01-NB3-B20497 & SFACT J214126.52+195850.3 & 325.36050 &  19.98063 & SFG &   0.14220 &  6563 & -0.658 $\pm$  0.103 & -0.004 $\pm$  0.050 & \nodata \\
 SFF01-NB3-B19399 & SFACT J214131.74+195847.3 & 325.38226 &  19.97982 & SFG &   0.13706 &  6563 & -0.816 $\pm$  0.101 &  0.321 $\pm$  0.021 & \nodata \\
 SFF01-NB2-D17902 & SFACT J214132.16+202146.6 & 325.38397 &  20.36294 & \nodata & \nodata & \nodata & \nodata & \nodata & \nodata \\
 SFF01-NB2-B19207 & SFACT J214132.90+193945.5 & 325.38708 &  19.66263 & SFG &   0.00350 &  6563 & (-1.212 $\pm$  0.151) & \nodata & \nodata \\
  \\
 SFF01-NB2-B19198 & SFACT J214132.93+193942.1 & 325.38721 &  19.66168 & HII &   0.00344 &  6563 & \nodata & \nodata & \nodata \\
 SFF01-NB1-B19076 & SFACT J214133.33+194226.7 & 325.38889 &  19.70742 & \nodata & \nodata & \nodata & \nodata & \nodata & \nodata \\
 SFF01-NB3-B18885 & SFACT J214134.29+194341.1 & 325.39285 &  19.72807 & ELG &   0.99895 &  3727 & \nodata & \nodata & \nodata \\
 SFF01-NB3-B18561 & SFACT J214135.87+194757.5 & 325.39948 &  19.79930 & ELG &   0.99635 &  3727 & \nodata & \nodata & \nodata \\
 SFF01-NB2-B18506 & SFACT J214136.06+195653.7 & 325.40027 &  19.94826 & ELG &   0.76173 &  3727 & \nodata & \nodata & \nodata \\
 SFF01-NB3-B18371 & SFACT J214136.92+193734.2 & 325.40381 &  19.62615 & \nodata & \nodata & \nodata & \nodata & \nodata & \nodata \\
 SFF01-NB3-B17245 & SFACT J214141.87+195707.7 & 325.42444 &  19.95213 & SFG &   0.13816 &  6563 & \nodata &  (0.746 $\pm$  0.158) & \nodata \\
 SFF01-NB3-D15415 & SFACT J214142.73+201252.3 & 325.42807 &  20.21452 & SFG &   0.13873 &  6563 & \nodata &  (0.617 $\pm$  0.151) & \nodata \\
 SFF01-NB2-D15191 & SFACT J214143.48+200448.4 & 325.43118 &  20.08011 & SFG &   0.31323 &  5007 & \nodata &  0.365 $\pm$  0.078 &  0.446 $\pm$  0.096 \\
 SFF01-NB1-B16317 & SFACT J214146.59+194328.3 & 325.44412 &  19.72453 & SFG &   0.39873 &  4959 & \nodata &  0.435 $\pm$  0.103 &  0.187 $\pm$  0.153 \\
 \\
 SFF01-NB3-B16011 & SFACT J214148.07+195758.9 & 325.45029 &  19.96636 & SFG &   0.14195 &  6563 & (-1.115 $\pm$  0.150) &  0.538 $\pm$  0.043 & \nodata \\
 SFF01-NB3-B15732 & SFACT J214149.54+194038.6 & 325.45642 &  19.67739 & \nodata & \nodata & \nodata & \nodata & \nodata & \nodata \\
 SFF01-NB2-B15722 & SFACT J214149.60+193848.3 & 325.45667 &  19.64676 & ELG &   0.76295 &  3727 & \nodata & \nodata & \nodata \\
 SFF01-NB1-D13076 & SFACT J214150.05+200728.2 & 325.45856 &  20.12449 & ELG &   0.85580 &  3727 & \nodata & \nodata & \nodata \\
 SFF01-NB1-D12776 & SFACT J214150.91+202341.9 & 325.46213 &  20.39497 & FD & \nodata & \nodata & \nodata & \nodata & \nodata \\
 SFF01-NB3-B14965 & SFACT J214153.12+195235.0 & 325.47134 &  19.87638 & SFG &   0.13832 &  6563 & (-0.966 $\pm$  0.148) &  0.293 $\pm$  0.109 & \nodata \\
 SFF01-NB3-B14799 & SFACT J214153.89+195227.9 & 325.47455 &  19.87441 & FD & \nodata & \nodata & \nodata & \nodata & \nodata \\
 SFF01-NB3-D11444 & SFACT J214155.02+201511.4 & 325.47925 &  20.25316 & SFG &   0.13740 &  6563 & (-1.050 $\pm$  0.146) &  0.678 $\pm$  0.091 & \nodata \\
 SFF01-NB2-B14205 & SFACT J214156.04+195310.0 & 325.48349 &  19.88612 & SFG &   0.32128 &  5007 & \nodata &  0.390 $\pm$  0.112 &  0.604 $\pm$  0.126 \\
 SFF01-NB1-D10796 & SFACT J214157.21+201512.0 & 325.48840 &  20.25334 & \nodata & \nodata & \nodata & \nodata & \nodata & \nodata \\
    \enddata
    \tablecomments{Table \ref{tab:specdata_sfactf01} is published in its entirety in the machine-readable format. A portion is shown here for guidance regarding its form and content.}
    \end{deluxetable*}

    \begin{deluxetable*}{ccccccccccccc}
    \tabletypesize{\scriptsize}
   \tablecaption{SFF10 Spectral Data Table \label{tab:specdata_sfactf10}}
    \tablehead{
    \\
     SFACT Object ID & SFACT Coordinate ID & $\alpha$(J2000) & $\delta$(J2000) & Type & z & Line & log([\ion{N}{2}]/H$\alpha$) & log([\ion{O}{3}]/H$\beta$) & log([\ion{O}{2}]/H$\beta$) \\
    & & degrees & degrees & & & &  \\  %$\frac{\text{erg}}{\text{s } \text{cm}^2}$
    (1) & (2) & (3) & (4) & (5) & (6) & (7) & (8) & (9) & (10) 
    }
  \startdata
 SFF10-NB3-D13755 & SFACT J014256.70+281615.3 &  25.73624 &  28.27093 & SFG &   0.49790 &  5007 & \nodata & \nodata & \nodata \\
 SFF10-NB3-D13569 & SFACT J014258.14+275740.4 &  25.74223 &  27.96124 & SFG &   0.48292 &  5007 & \nodata &  0.334 $\pm$  0.045 &  0.799 $\pm$  0.043 \\
 SFF10-NB2-B12883 & SFACT J014258.90+274309.1 &  25.74542 &  27.71919 & ELG &   0.77059 &  3727 & \nodata & \nodata & \nodata \\
 SFF10-NB3-B12772 & SFACT J014259.69+274052.3 &  25.74872 &  27.68121 & \nodata & \nodata & \nodata & \nodata & \nodata & \nodata \\
 SFF10-NB3-D13083 & SFACT J014300.87+280623.8 &  25.75362 &  28.10660 & SFG &   0.49586 &  5007 & \nodata &  0.664 $\pm$  0.060 & \nodata \\
 SFF10-NB1-B12579 & SFACT J014300.97+274122.1 &  25.75403 &  27.68948 & ELG &   0.86758 &  3727 & \nodata & \nodata & \nodata \\
 SFF10-NB3-B12471 & SFACT J014301.74+273928.5 &  25.75727 &  27.65791 & ELG &   1.00192 &  3727 & \nodata & \nodata & \nodata \\
 SFF10-NB1-D12909 & SFACT J014302.15+281152.9 &  25.75895 &  28.19802 & SFG &   0.39568 &  5007 & \nodata &  0.154 $\pm$  0.029 &  0.595 $\pm$  0.028 \\
 SFF10-NB3-B12244 & SFACT J014303.57+273655.2 &  25.76489 &  27.61533 & SFG &   0.48965 &  5007 & \nodata & \nodata & \nodata \\
 SFF10-NB2-B12225 & SFACT J014303.75+273744.4 &  25.76562 &  27.62900 & FD & \nodata & \nodata & \nodata & \nodata & \nodata \\
 \\
 SFF10-NB3-B12131 & SFACT J014304.20+275025.1 &  25.76750 &  27.84031 & SFG &   0.13419 &  6563 & -0.803 $\pm$  0.101 &  0.180 $\pm$  0.035 & \nodata \\
 SFF10-NB3-B12081 & SFACT J014304.52+275030.9 &  25.76884 &  27.84191 & SFG &   0.13458 &  6563 & -0.880 $\pm$  0.085 &  0.061 $\pm$  0.057 & \nodata \\
 SFF10-NB1-B12096 & SFACT J014304.64+273418.2 &  25.76935 &  27.57173 & FD & \nodata & \nodata & \nodata & \nodata & \nodata \\
 SFF10-NB3-D12508 & SFACT J014305.18+275632.9 &  25.77158 &  27.94246 & SFG &   0.13950 &  6563 & \nodata & \nodata & \nodata \\
 SFF10-NB3-B11870 & SFACT J014306.06+274431.6 &  25.77523 &  27.74211 & ELG &   1.00985 &  3727 & \nodata & \nodata & \nodata \\
 SFF10-NB3-B11533 & SFACT J014307.68+275244.6 &  25.78199 &  27.87906 & ELG &   0.98897 &  3727 & \nodata & \nodata & \nodata \\
 SFF10-NB3-D12024 & SFACT J014308.50+281103.9 &  25.78543 &  28.18441 & ELG &   1.01272 &  3727 & \nodata & \nodata & \nodata \\
 SFF10-NB2-D11997 & SFACT J014308.81+281226.9 &  25.78671 &  28.20746 & QSO &   2.48155 &  1908 & \nodata & \nodata & \nodata \\
 SFF10-NB3-D11938 & SFACT J014309.34+280305.7 &  25.78890 &  28.05159 & SFG &   0.13572 &  6563 & -1.542 $\pm$  0.097 &  0.640 $\pm$  0.019 & \nodata \\
 SFF10-NB3-D11925 & SFACT J014309.51+275434.0 &  25.78963 &  27.90944 & FD & \nodata & \nodata & \nodata & \nodata & \nodata \\
 \\
 SFF10-NB3-B11195 & SFACT J014309.97+273813.9 &  25.79156 &  27.63718 & ELG &   1.00344 &  3727 & \nodata & \nodata & \nodata \\
 SFF10-NB3-D11772 & SFACT J014310.63+280641.6 &  25.79429 &  28.11155 & SFG &   0.49388 &  5007 & \nodata &  0.814 $\pm$  0.067 & -0.196 $\pm$  0.122 \\
 SFF10-NB1-B10986 & SFACT J014311.40+273300.7 &  25.79748 &  27.55019 & \nodata & \nodata & \nodata & \nodata & \nodata & \nodata \\
 SFF10-NB1-B10674 & SFACT J014313.32+274124.6 &  25.80549 &  27.69016 & SFG &   0.06418 &  6563 & -1.475 $\pm$  0.088 &  0.543 $\pm$  0.014 & \nodata \\
 SFF10-NB2-B10675 & SFACT J014313.39+273356.3 &  25.80581 &  27.56564 & \nodata & \nodata & \nodata & \nodata & \nodata & \nodata \\
 SFF10-NB1-D11121 & SFACT J014315.23+280125.9 &  25.81347 &  28.02386 & ELG &   0.87527 &  3727 & \nodata & \nodata & \nodata \\
 SFF10-NB3-D11032 & SFACT J014315.47+281521.9 &  25.81444 &  28.25607 & SFG &   0.49829 &  5007 & \nodata &  0.739 $\pm$  0.129 &  0.835 $\pm$  0.129 \\
 SFF10-NB3-D10890 & SFACT J014316.09+275845.4 &  25.81705 &  27.97927 & HII &   0.13445 &  6563 & -0.653 $\pm$  0.054 &  0.097 $\pm$  0.040 & \nodata \\
 SFF10-NB3-D10873 & SFACT J014316.14+275843.3 &  25.81726 &  27.97869 & SFG &   0.13479 &  6563 & -0.519 $\pm$  0.042 & -0.458 $\pm$  0.105 & \nodata \\
 SFF10-NB3-B10141 & SFACT J014316.41+274430.5 &  25.81836 &  27.74181 & ELG &   0.99417 &  3727 & \nodata & \nodata & \nodata \\
\enddata
    \tablecomments{Table \ref{tab:specdata_sfactf10} is published in its entirety in the machine-readable format. A portion is shown here for guidance regarding its form and content.}
    \end{deluxetable*}

    \begin{deluxetable*}{ccccccccccccc}
    \tabletypesize{\scriptsize}
   \tablecaption{SFF15 Spectral Data Table \label{tab:specdata_sfactf15}}
    \tablehead{
    \\
     SFACT Object ID & SFACT Coordinate ID & $\alpha$(J2000) & $\delta$(J2000) & Type & z & Line & log([\ion{N}{2}]/H$\alpha$) & log([\ion{O}{3}]/H$\beta$) & log([\ion{O}{2}]/H$\beta$) \\
    & & degrees & degrees & & & &  \\  %$\frac{\text{erg}}{\text{s } \text{cm}^2}$
    (1) & (2) & (3) & (4) & (5) & (6) & (7) & (8) & (9) & (10) 
    }
    \startdata
 SFF15-NB3-B14284 & SFACT J023730.56+274425.1 &  39.37733 &  27.74032 & SFG &   0.13180 &  6563 & (-0.876 $\pm$  0.147) &  0.552 $\pm$  0.110 & \nodata \\
 SFF15-NB1-B14251 & SFACT J023730.82+274059.9 &  39.37840 &  27.68330 & ELG &   0.86741 &  3727 & \nodata & \nodata & \nodata \\
 SFF15-NB2-D24348 & SFACT J023731.19+281026.8 &  39.37994 &  28.17412 & SFG &   0.31444 &  5007 & \nodata &  0.251 $\pm$  0.104 &  0.446 $\pm$  0.115 \\
 SFF15-NB2-B14131 & SFACT J023731.68+272845.4 &  39.38200 &  27.47929 & \nodata & \nodata & \nodata & \nodata & \nodata & \nodata \\
 SFF15-NB3-B14046 & SFACT J023731.96+275052.4 &  39.38317 &  27.84789 & SFG &   0.54159 &  4861 & \nodata & \nodata & \nodata \\
 SFF15-NB2-D22777 & SFACT J023733.00+281011.2 &  39.38748 &  28.16978 & SFG &   0.31010 &  5007 & \nodata &  0.624 $\pm$  0.031 &  0.589 $\pm$  0.035 \\
 SFF15-NB2-B13724 & SFACT J023734.14+274128.0 &  39.39226 &  27.69111 & SFG &   0.31705 &  5007 & \nodata &  0.660 $\pm$  0.071 &  0.505 $\pm$  0.095 \\
 SFF15-NB2-B13721 & SFACT J023734.15+274129.1 &  39.39229 &  27.69142 & SFG &   0.31786 &  5007 & \nodata &  (0.241 $\pm$  0.164) &  (0.829 $\pm$  0.156) \\
 SFF15-NB2-D22292 & SFACT J023734.35+281003.1 &  39.39310 &  28.16751 & SFG &   0.32192 &  5007 & \nodata &  0.325 $\pm$  0.083 &  0.617 $\pm$  0.089 \\
 SFF15-NB2-D22224 & SFACT J023734.52+280514.9 &  39.39383 &  28.08749 & SFG &   0.30904 &  5007 & \nodata &  (0.776 $\pm$  0.150) & \nodata \\
 \\
 SFF15-NB3-D22277 & SFACT J023734.53+275653.7 &  39.39389 &  27.94825 & SFG &   0.48883 &  5007 & \nodata & \nodata & \nodata \\
 SFF15-NB3-B13580 & SFACT J023735.23+274121.9 &  39.39680 &  27.68941 & \nodata & \nodata & \nodata & \nodata & \nodata & \nodata \\
 SFF15-NB2-D20719 & SFACT J023738.93+280530.2 &  39.41222 &  28.09174 & \nodata & \nodata & \nodata & \nodata & \nodata & \nodata \\
 SFF15-NB1-B12855 & SFACT J023740.25+274942.1 &  39.41772 &  27.82837 & ELG &   0.85732 &  3727 & \nodata & \nodata & \nodata \\
 SFF15-NB2-B12874 & SFACT J023740.34+272939.9 &  39.41807 &  27.49442 & SFG &   0.32207 &  5007 & \nodata &  0.669 $\pm$  0.054 &  0.010 $\pm$  0.108 \\
 SFF15-NB2-B12729 & SFACT J023741.60+272959.4 &  39.42332 &  27.49985 & SFG &   0.32167 &  5007 & \nodata &  0.813 $\pm$  0.096 &  0.489 $\pm$  0.117 \\
 SFF15-NB2-D19839 & SFACT J023741.93+281035.0 &  39.42469 &  28.17639 & SFG &   0.33183 &  4959 & \nodata &  0.650 $\pm$  0.055 &  0.461 $\pm$  0.070 \\
 SFF15-NB1-B12675 & SFACT J023742.16+272938.9 &  39.42568 &  27.49414 & \nodata & \nodata & \nodata & \nodata & \nodata & \nodata \\
 SFF15-NB3-B12608 & SFACT J023742.63+274616.0 &  39.42762 &  27.77110 & SFG &   0.13623 &  6563 & \nodata &  (0.894 $\pm$  0.149) & \nodata \\
 SFF15-NB3-B12427 & SFACT J023744.04+274533.7 &  39.43350 &  27.75936 & SFG &   0.13181 &  6563 & (-0.866 $\pm$  0.154) &  0.143 $\pm$  0.111 & \nodata \\
 \\
 SFF15-NB3-D18788 & SFACT J023744.26+281330.5 &  39.43440 &  28.22513 & \nodata & \nodata & \nodata & \nodata & \nodata & \nodata \\
 SFF15-NB2-B12371 & SFACT J023744.60+273441.2 &  39.43585 &  27.57810 & ELG &   0.77650 &  3727 & \nodata & \nodata & \nodata \\
 SFF15-NB1-D18539 & SFACT J023744.78+281340.3 &  39.43659 &  28.22785 & \nodata & \nodata & \nodata & \nodata & \nodata & \nodata \\
 SFF15-NB3-B12212 & SFACT J023745.92+273918.5 &  39.44135 &  27.65515 & SFG &   0.48206 &  5007 & \nodata &  0.584 $\pm$  0.054 &  0.658 $\pm$  0.053 \\
 SFF15-NB3-D17352 & SFACT J023746.87+280421.5 &  39.44530 &  28.07263 & FD & \nodata & \nodata & \nodata & \nodata & \nodata \\
 SFF15-NB3-B11965 & SFACT J023747.82+274851.5 &  39.44925 &  27.81431 & FD & \nodata & \nodata & \nodata & \nodata & \nodata \\
 SFF15-NB1-D15218 & SFACT J023748.24+281511.6 &  39.45099 &  28.25321 & \nodata & \nodata & \nodata & \nodata & \nodata & \nodata \\
 SFF15-NB1-D14190 & SFACT J023749.21+280902.6 &  39.45504 &  28.15072 & ELG &   0.79227 &  3869 & \nodata & \nodata & \nodata \\
 SFF15-NB2-B11684 & SFACT J023750.76+272858.2 &  39.46150 &  27.48282 & SFG &   0.30978 &  5007 & \nodata &  0.764 $\pm$  0.119 & \nodata \\
 SFF15-NB3-D13288 & SFACT J023751.99+275346.3 &  39.46664 &  27.89621 & ELG &   0.99454 &  3727 & \nodata & \nodata & \nodata \\
\enddata
    \tablecomments{Table \ref{tab:specdata_sfactf15} is published in its entirety in the machine-readable format. A portion is shown here for guidance regarding its form and content.}
    \end{deluxetable*}

%    \enddata
%    \tablecomments{Here N2\HA, O3\HB, and O2\HB\ represent $\log_{10}$ values for the individual line ratios. Table \ref{tab:specdata_sfactf15} is published in its entirety in the machine-readable format. A portion is shown here for guidance regarding its form and content.}\end{deluxetable*}

%************************************************************************************************************************

\section{Results: Presentation of Spectral Data from the Pilot Study} \label{sec:results}
%section summary
    To date 453 out of the 533 objects in the three pilot-study fields have been observed spectroscopically and 415 of the observed 453 are confirmed emission-line objects. That means 91.6\% of our sources observed have an emission line within the narrow-band filter they were detected in and are determined to be true emission-line sources. 
    
In this section, we present the results from the spectroscopic observations of the SFACT pilot-study fields.  In Section \ref{sec:specdatatab} we tabulate the relevant spectral data for all three pilot-study fields. Next, Section \ref{sec:spectra} provides illustrative examples of the spectra for a number of SFACT objects.  Section \ref{sec:properties}  presents an overview of the different emission lines detected in the pilot study as well as some of the properties of the ELGs.  This includes a discussion of the redshift distribution of the sample and a presentation of the emission-line diagnostic diagrams derived from the spectral data.

\subsection{SFACT Spectral Data Tables} \label{sec:specdatatab}

    Tables \ref{tab:specdata_sfactf01}, \ref{tab:specdata_sfactf10}, and \ref{tab:specdata_sfactf15} present the spectral data for the SFACT objects in the SFF01, SFF10, and SFF15 fields, respectively. 
    
    Column (1) in each table lists the SFACT object identifier. This is a unique ID for each object and contains three parts. The first part represents the field where the object is located (e.g., SFF01), while the second designates the filter within which the source was detected. The third contains information on which quadrant of the field the object is located in (A, B, C, or D) followed by a number that represents the order in which the image processing software finds the object in the field.   We use the SFACT object ID to refer to specific sources throughout the remainder of this paper (e.g., spectral plots in Section \ref{sec:examplespec}).  Column (2) provides an alternate coordinate-based designation, using IAU-approved nomenclature.
    
      Columns (3) and (4) list the RA and Dec of each source (J2000).  Each table is sorted in ascending RA order.  The SFACT coordinates are derived from an astrometric solution applied to the survey imaging data based on the Gaia database (\citealt{2016A&A...595A...1G, 2021A&A...649A...1G}).   Comparison of the coordinates of stars found in the SFACT images with those cataloged in the SDSS shows that there is little or no systematic offsets between the two sets of coordinates (mean $\Delta\alpha$ and $\Delta\delta$ $\leq$ 0.05 arcsec for each field), and that the RMS scatter for individual stars is $\sim$0.15--0.20 arcsec.
    
    Column (5) indicates the activity type of each source detected by the survey, where we adopt standard notations for the various classes of emission-line objects.   The activity type is derived by visual inspection of the spectra, supplemented by use of line diagnostic diagrams (see \S \ref{sec:emissionlinediagnosticdiagrams}).  The vast majority of objects detected by SFACT are star-forming galaxies, which are labeled SFG in the tables.  We make no attempt to differentiate between different classes of SFGs here (e.g., Starburst Nucleus galaxies, blue compact dwarfs, Green Pea galaxies) since that would typically require additional information not available from the spectral data alone.   However, objects specified as \ion{H}{2} regions is nearby disk galaxies (see SFACT1) are labeled as HII in column (5).   Sy1 and Sy2 labels indicate that the listed object is either a Seyfert 1 or Seyfert 2 active galactic nucleus (AGN).  LIN indicates the object is a low-ionization nuclear emission region (LINER; \citealt{1980A&A....87..152H}), and QSO indicates the object is a quasar.  Objects that are clear detections of ELGs with an emission line in the appropriate filter, but have an uncertain classification at the time the lines are measured, are marked with ELG as the default designation.   The latter classification is currently applied to the vast majority of the [\ion{O}{2}]-detected SFACT objects, since we lack spectral information from other diagnostic lines that would allow us to make a more definitive designation.
    
    It is inevitable that false detections will creep into a survey like SFACT.  These false detections have their type labeled in two ways. Objects that have no obvious emission lines in their spectra, or that have lines that are not located in the relevant survey filter, are simply labeled as FD (for false detection).  Objects found to be stars based on their spectra are labeled with Star in this column.  Some of these are stars with an emission line in the relevant filter (always NB2); these are specified in the table notes.  An example of such an object is shown in Figure 11 of SFACT1.  Finally, objects without any data in column (5) are objects that have yet to be observed spectroscopically. 
    
    Column (6) displays the redshift of each object.  The characteristic uncertainties in our redshift measurements are 0.00003 to 0.00005, based on the RMS scatter of the redshifts measured from individual lines in a given spectrum (see  \S \ref{sec:wralf}).   The emission line responsible for the detection of each SFACT object is listed in column (7).  That is, the emission line indicated is the one responsible for most or all of the excess emission present in the relevant NB survey filter.  The majority of the SFACT galaxies are detected via either H$\alpha$ $\lambda$6563, [\ion{O}{3}]$\lambda$5007 or [\ion{O}{2}]$\lambda$3727, but small numbers of objects are detected via other lines, such as [\ion{O}{3}]$\lambda$4959, [\ion{S}{2}]$\lambda\lambda$6717,6731, [\ion{N}{3}]$\lambda$3869, and H$\beta$ $\lambda$4861.  SFACT QSOs are typically detected via one of the stronger UV lines: \ion{Mg}{2} $\lambda$2798, \ion{C}{3}] $\lambda$1908, \ion{C}{4} $\lambda$1549, or Ly$\alpha$ $\lambda$1215.
    
     Finally, columns (8) through (10) show the $\log_{10}$ values for the \NII/\HA, \OIII/\HB, and \OII/\HB\ line ratios and their formal errors. For line ratios were one or both lines are Category 2 re-examination measurements (described in \S 4.2), the uncertainties in the line ratios will be larger.  We denote these cases in the data tables by enclosing the line ratios in parentheses. The line ratios are corrected for reddening using the Balmer decrement method \citep[e.g., ][]{2006agna.book.....O} when the relevant Balmer lines are available in our spectra.  100\% of the H$\alpha$-detected SFACT objects have the necessary H$\alpha$ and H$\beta$ lines for this correction.  The [\ion{O}{3}]-detected sources are corrected using the H$\beta$ and H$\gamma$ lines, both of which are present in only $\sim$25\% of the spectra.

A given line ratio will only be measured if both of the relevant emission lines are within the observed spectral wavelength range (4760$-$7580~\AA) {\it and} are measured by our software.  Because SFACT discovers galaxies located within discrete redshift windows and we have used a fixed spectral coverage for our follow-up spectra, objects detected via a given emission line (column (7)) will always have the same set of emission-line ratios available.    For example, H$\alpha$-detected galaxies will always have the \NII/\HA\ and \OIII/\HB\ ratios but never \OII/\HB.  Similarly, [\ion{O}{3}]-detected galaxies will possess the \OIII/\HB\ and \OII/\HB\ ratios in our tables, but never \NII/\HA.   Galaxies detected via their [\ion{O}{2}] emission will have no line ratios listed.
 
    For the reasons specified in \S~\ref{sec:limitations} our survey data tables do not include the individual line EWs or fluxes.  While the majority of our EWs are reliable, the nature of the sky-subtraction uncertainty is such that it is impossible to know with confidence which objects possess less-robust values which should be ignored.   For the purposes of calculating relevant physical quantities such as star-formation rates (SFR) the preferred methodology would be to use the emission-line flux measured from the NB images, since the latter will typically be a more robust measurement and will include all of the flux from each source.  Furthermore, the fluxes of lines other than the one detected in the NB imaging survey can, in most cases,  be scaled up using the relative line ratios from the spectra.

%************************************************************************************************************************

\subsection{Example SFACT Spectra} \label{sec:spectra}

\subsubsection{Spectra of example objects illustrated in SFACT2} \label{sec:examplespec}

\begin{figure}[t]
    \centering
    \includegraphics[width=\columnwidth,keepaspectratio]{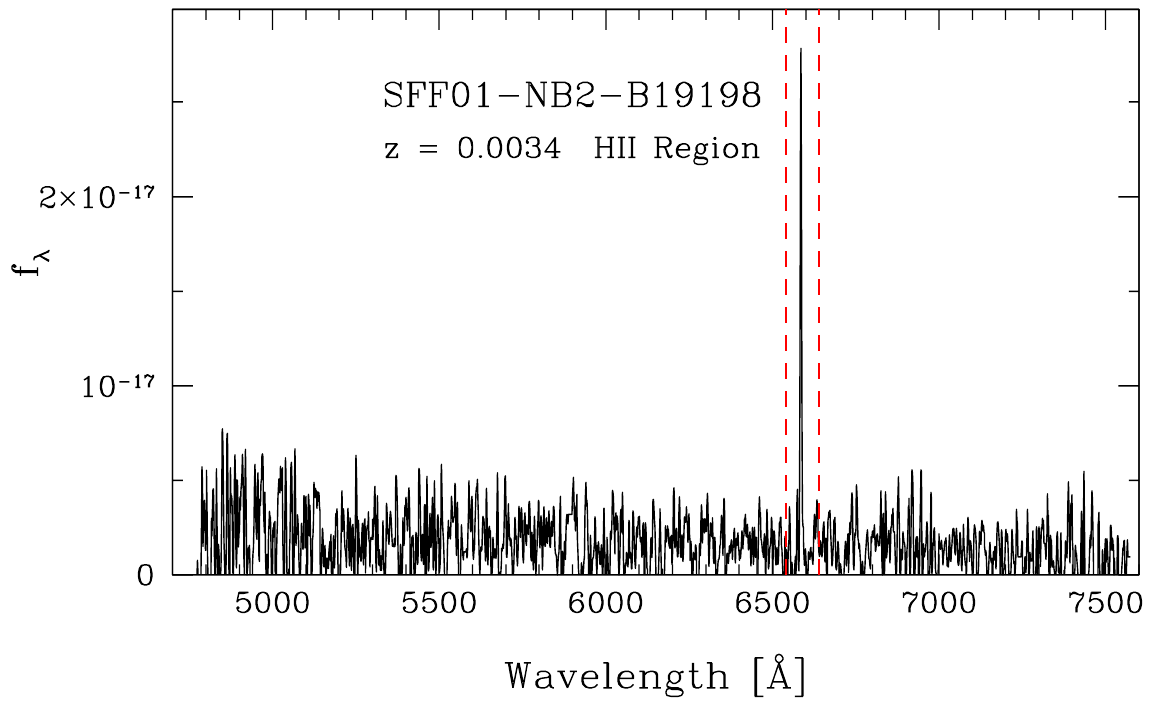}
    \put(-200,125){\large (a)}
    
    \includegraphics[width=\columnwidth,keepaspectratio]{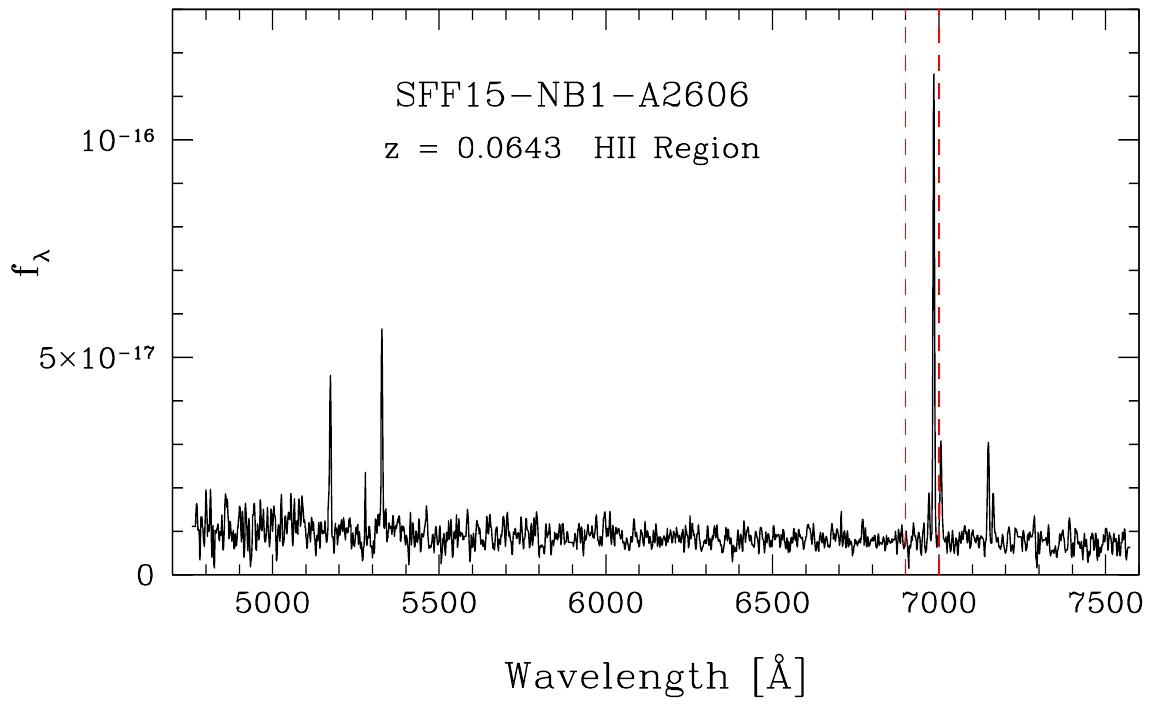}
    \put(-200,125){\large (b)}
    \put(-43,130){\scriptsize H$\alpha$+[N II]}
    \put(-41,62){\scriptsize [S II]}
    \put(-186,70){\scriptsize H$\beta$}
    \put(-172,86){\scriptsize [O III]}

    \includegraphics[width=\columnwidth,keepaspectratio]{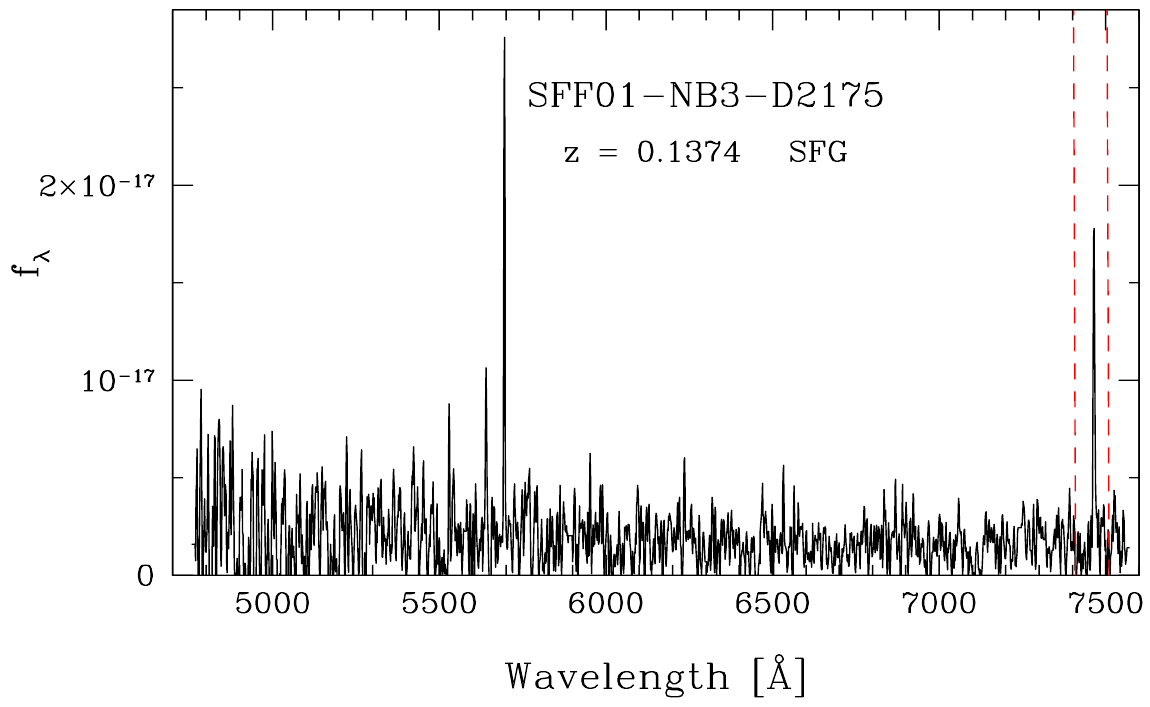}
    \put(-200,125){\large (c)}
    
    \caption{Spectra of three objects detected by their \HA\ lines. The red-dashed vertical lines denote the wavelength range covered by the narrow-band filter in which the source was detected.  Key emission lines are labelled in the middle panel.  The flux scales on the y-axes are in units of erg s$^{-1}$ cm$^{-2}$ \AA$^{-1}$.
    (a): Detection in NB2 of an outlying \HII\ region in a dwarf irregular galaxy. This is the lowest redshift source in the SFACT pilot-study fields.
    (b): An \HII\ region detected in a spiral galaxy with the NB1 filter. 
    (c): A star-forming galaxy detected in the NB3 filter. Its \OIII$\lambda$5007/\HB\ ratio indicates that this is a low-metallicity system.
    \label{fig:ha}}
    \end{figure}

\begin{figure}[t]
    \centering
    \includegraphics[width=\columnwidth,keepaspectratio]{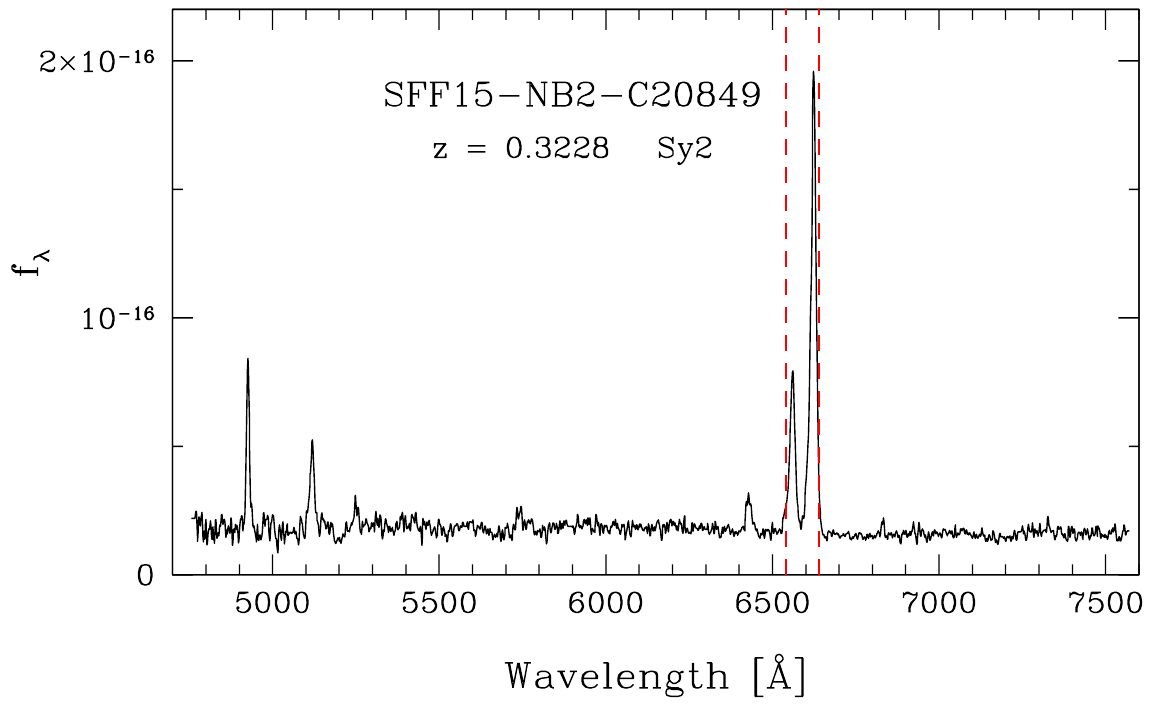}
    \put(-200,125){\large (a)}
    \put(-92,51){\scriptsize H$\beta$}
    \put(-67,120){\scriptsize [O III]}
    \put(-140,47){\scriptsize H$\gamma$}
    \put(-186,61){\scriptsize [Ne III]}
    \put(-198,80){\scriptsize [O II]}

    \includegraphics[width=\columnwidth,keepaspectratio]{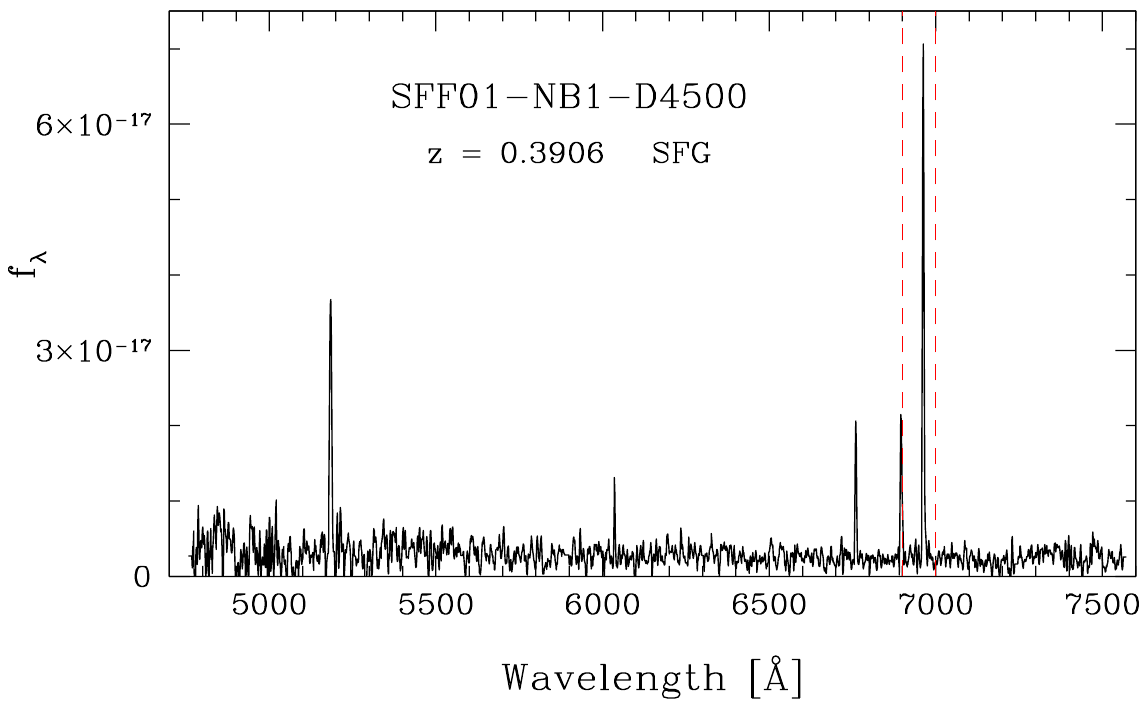}
    \put(-200,125){\large (b)}
    
    \includegraphics[width=\columnwidth,keepaspectratio]{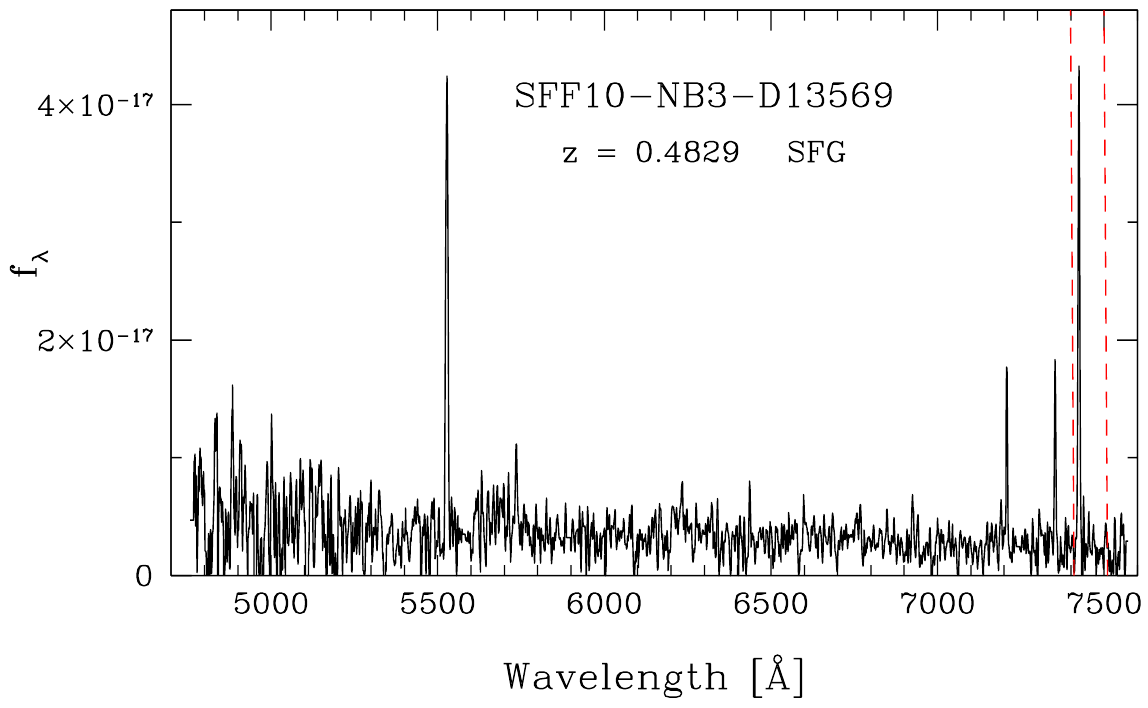}
    \put(-200,125){\large (c)}
    
    \caption{Spectra of three objects detected by their \OIII$\lambda$5007 lines. The red-dashed vertical lines denote the wavelength range covered by the narrow-band filter in which the source was detected.  Key emission lines are labelled in the top panel.  The flux scales on the y-axes are in units of erg s$^{-1}$ cm$^{-2}$ \AA$^{-1}$.
    (a): Seyfert 2 galaxy detected in the NB2 filter. 
    (b): Star-forming galaxy detected in the NB1 filter.
    (c): Star-forming system detected in the NB3 filter.
    \label{fig:OIII}}
    \end{figure}

\begin{figure}[t]
    \centering
    \includegraphics[width=\columnwidth,keepaspectratio]{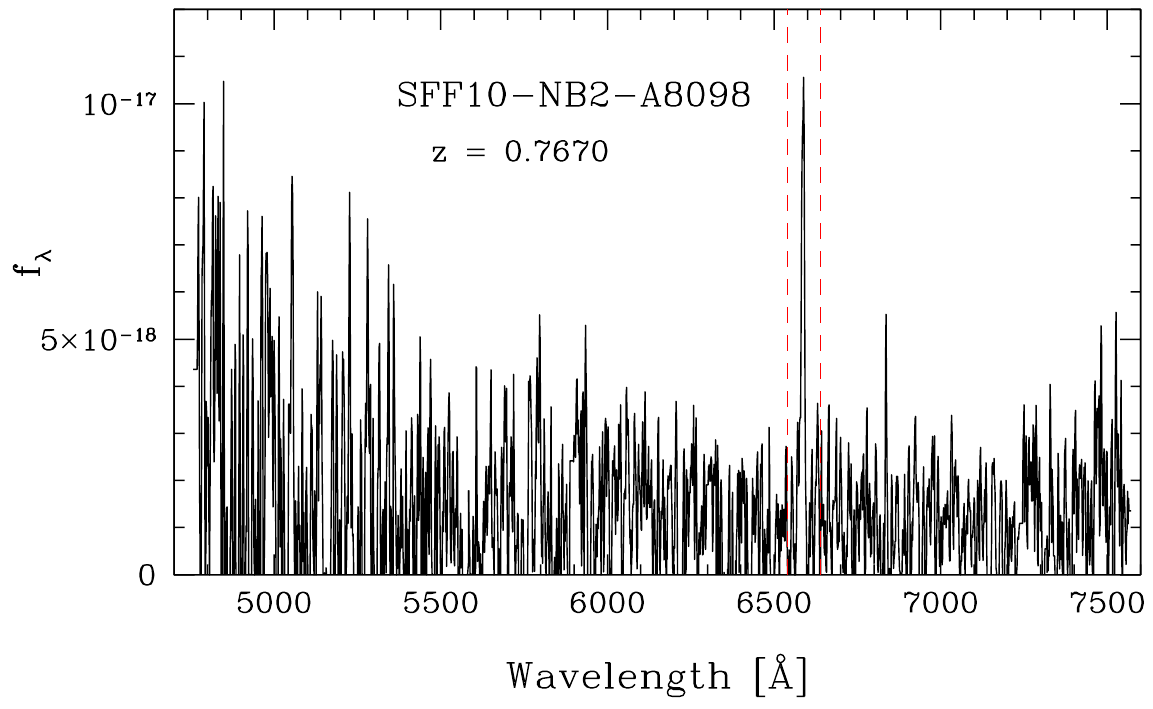}
    \put(-200,125){\large (a)}

    \includegraphics[width=\columnwidth,keepaspectratio]{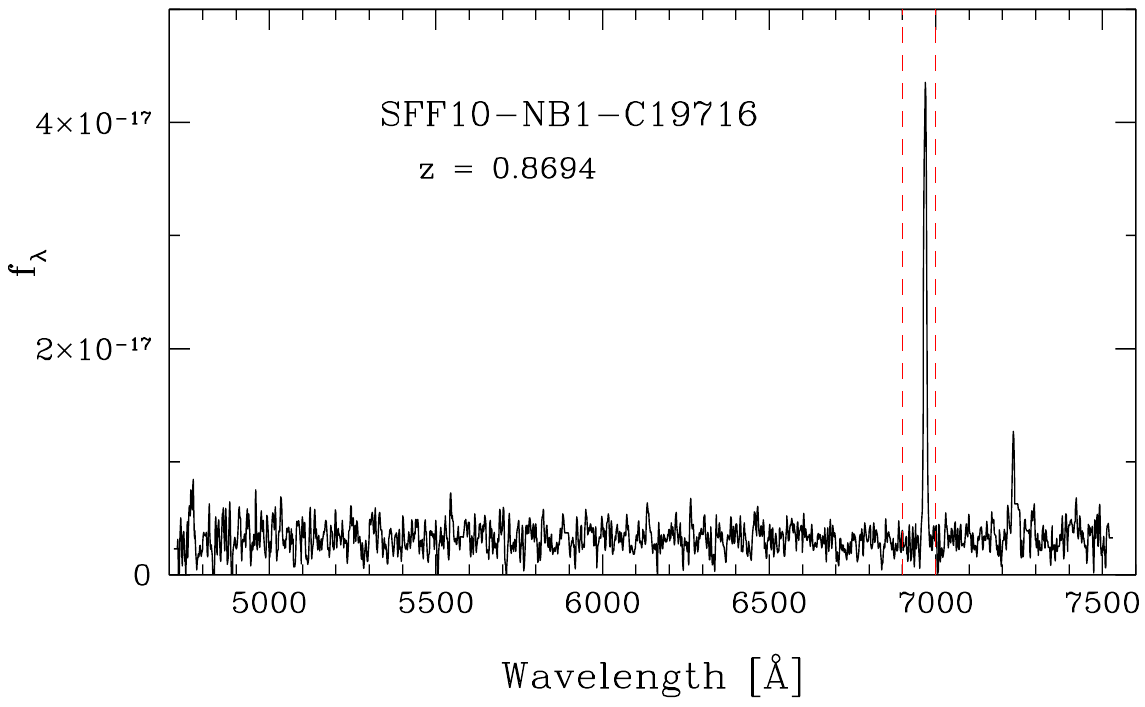}
    \put(-200,125){\large (b)}
    \put(-39,63){\scriptsize [Ne III]}
    \put(-41,120){\scriptsize [O II]}
    
    \includegraphics[width=\columnwidth,keepaspectratio]{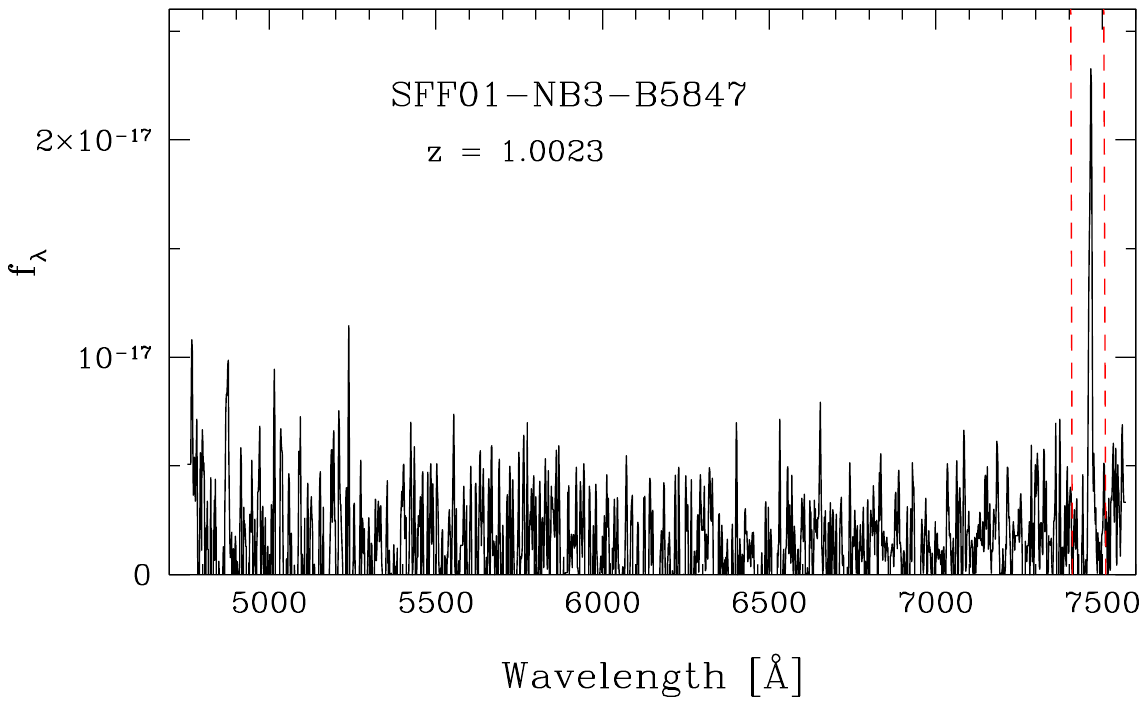}
    \put(-200,125){\large (c)}
    
    \caption{Spectra of three objects detected by their \OII$\lambda$3727 doublet. The red-dashed vertical lines denote the wavelength range covered by the narrow-band filter in which the source was detected. Key emission lines are labelled in the middle panel.  The flux scales on the y-axes are in units of erg s$^{-1}$ cm$^{-2}$ \AA$^{-1}$.
    (a): An NB2-detected galaxy, included to show the faint nature of some of our sources. 
    (b): A galaxy detected by the NB1 filter. \NeIII$\lambda$3869 is visible in both this spectrum and in panel (a).
    (c): An NB3-detected source,  showing the limited information present in the spectra of the NB3 \OII\ detections. 
    \label{fig:OII}}
    \end{figure}

\begin{figure}[t]
    \centering
    \includegraphics[width=\columnwidth,keepaspectratio]{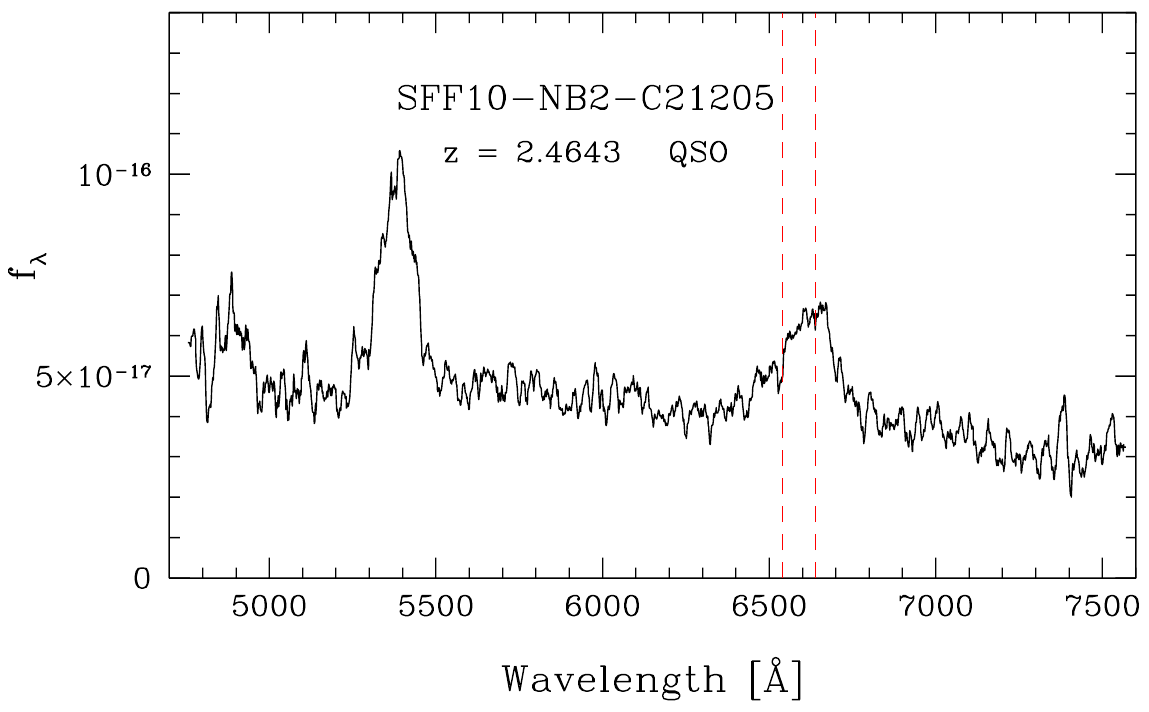}
    \put(-60,83){\scriptsize C III]}
    \put(-185,110){\scriptsize C IV}
    
    \caption{Spectrum of an SFACT-detected QSO. The red-dashed vertical lines denote the wavelength range covered by the narrow-band filter in which the source was detected.  Key emission lines are labelled.  The flux scale on the y-axis is in units of erg s$^{-1}$ cm$^{-2}$ \AA$^{-1}$.  This QSO was detected by its C~{\footnotesize III}]$\lambda$1908 line. 
    \label{fig:Quasar}}
    \end{figure}

    SFACT2 presents example images taken with both the broadband and narrowband filters used in our survey (Figures 3-6 in that paper).   These images include objects detected in each of the three NB filters and with each of the three primary emission lines (\HA, \OIII, and \OII) used to detect galaxies in the survey.  In addition, one SFACT QSO is also displayed.  The objects selected to illustrate the survey detections were also picked to show the range to emission-line fluxes detected in the NB filters.  Figures \ref{fig:ha} through \ref{fig:Quasar} in the current paper present  spectra of these same objects.  The object's SFACT identifier, redshift, and activity class are labeled on the plot. The red-dashed vertical lines denote the wavelength range covered by the narrow-band filter in which the source was detected. 
    
    %OII
    
    %Ha
    Figure \ref{fig:ha} shows some of the \HA-detected sources in the pilot study.  %Panel (a) and (b) in this figure show two \HII\ regions. 
The object in panel (a) is an \HII\ region in a dwarf irregular galaxy with a redshift of z~$= 0.0034$. It is the only \HA-detected galaxy in the NB2 filter in the pilot-study fields and is the lowest redshift source in the pilot study.  The NB2 filter probes a redshift range of $-$0.002 to 0.011 for the \HA\ line, resulting in a very small volume being surveyed. Hence, it is no surprise that the pilot study only contains one object that is an NB2 \HA\ detection (see \citetalias{Paper1}).   Panel (b) shows the spectrum of an  \HII\ region located in a spiral galaxy that was detected in NB1. It belongs to a fairly metal rich system, based on the observed ratios of  [\ion{N}{2}]/H$\alpha$ and [\ion{O}{3}]/H$\beta$ (see Figure~\ref{fig:BPT}).  This spectrum gives a good example of the additional lines we can detect when an object is selected by its \HA\ line: the \SII\ doublet, \NII$\lambda\lambda$6583, 6548, \OIII$\lambda\lambda 5007,4959$, and \HB\ lines are clearly visible.  Finally, panel (c) shows a low-metallicity star-forming system detected by the NB3 filter. It has a \textit{g}-band magnitude of 22.4, from which its derived absolute magnitude is calculated to be M$_g = -16.7$ (i.e., comparable to the luminosity of the SMC, but at a distance of $\sim$650 Mpc).
    
    %OIII
    In Figure \ref{fig:OIII}, we present three example spectra of \OIII-detected sources. Panel (a) is a Seyfert 2 galaxy detected in the NB2 filter. Its emission lines are clearly broader than the lines seen in the two star-forming galaxies shown in panels (b) and (c), and its line ratios are indicative of a non-stellar ionizing source (e.g., [\ion{O}{3}]/H$\beta$ $>$ 10). It has an absolute magnitude of M$_g$ = $-$19.9. The second and third spectra displayed in this figure are NB1 and NB3 detections. They each show star-forming systems and have M$_g$ = $-$19.1 and $-$19.2, respectively. Note that our \OIII-detected spectra usually contain additional lines like \OIII$\lambda$4959, \HB, \HG, and the \OII\ doublet.
    
    %OII
    Example spectra of \OII-detected sources are shown in Figure \ref{fig:OII}. As is the case with most of the \OII-detected galaxies, it is difficult to say much about the nature of these sources due to the limited number of lines detected in the survey's wavelength range. However, the \OII\ doublet is usually sufficiently resolved in our spectra so that the two lines that make up the doublet can be distinguished from each other. Additional lines, such as \NeIII$\lambda$3869, are also sometimes present in the \OII-detected spectra. \NeIII\ is visible in both panels (a) and (b) of the figure, though it is redshifted out of the survey's wavelength range in panel (c), the displayed NB3 detection.  If we wish to gain additional information about the spectral characteristics of these objects, such as their [\ion{N}{2}]/H$\alpha$ and [\ion{O}{3}]/H$\beta$ ratios, follow up observations redward of our current wavelength coverage will need to be carried out.
    
    The \OII-detected sources are among the faintest in apparent magnitude in the SFACT survey due to their distance. The spectrum in panel (a) of Figure \ref{fig:OII} comes from a source with a \textit{g}-band magnitude of 23.77 which is fainter than the 23.15 median of the sources in the pilot study (see \citetalias{Paper1}). This is not the faintest object in the SFACT survey but it is the faintest object presented in this series of examples. It has an absolute magnitude of M$_g$ = $-$19.7 while the objects presented in (b) and (c) have absolute magnitudes M$_g$ of $-20.8$ and $-21.1$. All three of these sources are intrinsically luminous, which makes sense given that the survey was able to detect them at such large redshifts. 
    
    %Quasar
     Figure \ref{fig:Quasar} shows the spectrum of a QSO detected in SFACT. The survey image data for this object is shown in Figure 6 of \citetalias{Paper2}. It is detected in the NB2 filter by its C~{\footnotesize III}]$\lambda$1908 line, and C~{\footnotesize IV}$\lambda$1549 is also seen in the spectrum. It has a redshift of z~$ = 2.46$ and an absolute magnitude of M$_g = -25.7$. This is a very luminous object with very broad emission lines, which is consistent with expectations for QSOs.

\subsubsection{Spectra of ELGs detected via non-standard lines} \label{sec:additionalspec}

    \begin{figure*}[t]
    \centering
    \includegraphics[width=0.49\textwidth,keepaspectratio]{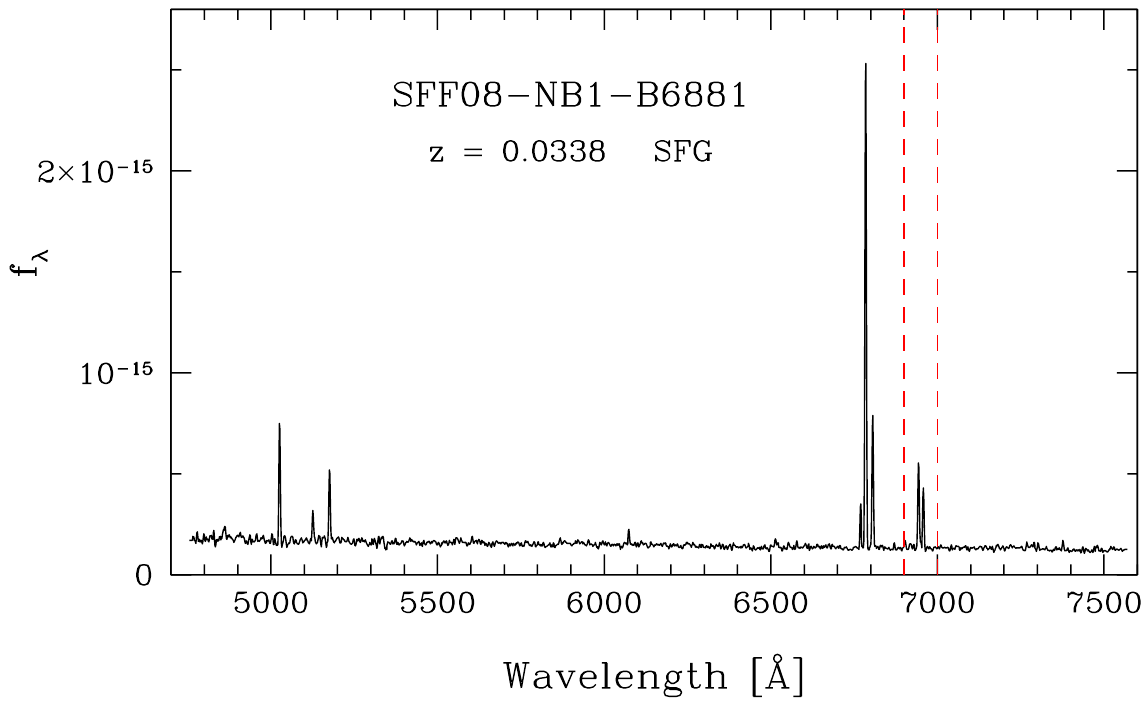} \includegraphics[width=0.49\textwidth,keepaspectratio]{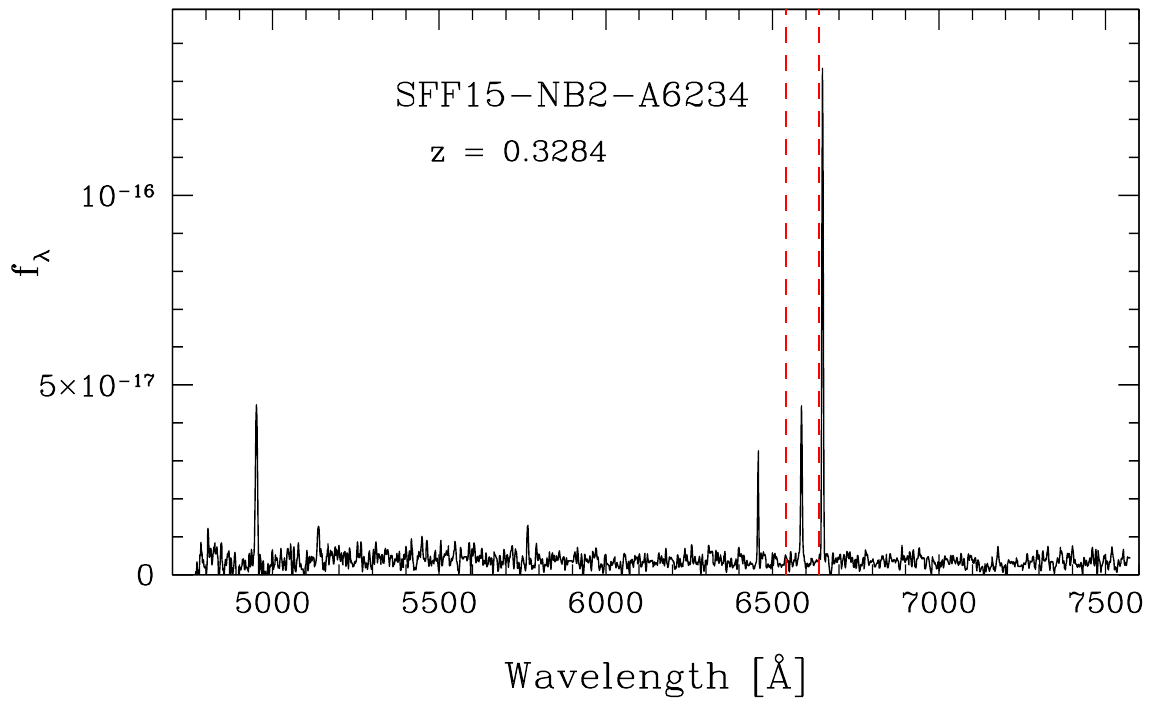}
    \put(-455,130){\large (a)}
    \put(-355,95){\scriptsize H$\alpha$+[N II]}
    \put(-299,55){\scriptsize [S II]}
    \put(-442,55){\scriptsize [O III]}
    \put(-450,65){\scriptsize H$\beta$}
    \put(-205,130){\large (b)}
    \put(-70,120){\scriptsize [O III]}
    \put(-93,60){\scriptsize H$\beta$}
    \put(-205,72){\scriptsize [O II]}

    \includegraphics[width=0.49\textwidth,keepaspectratio]{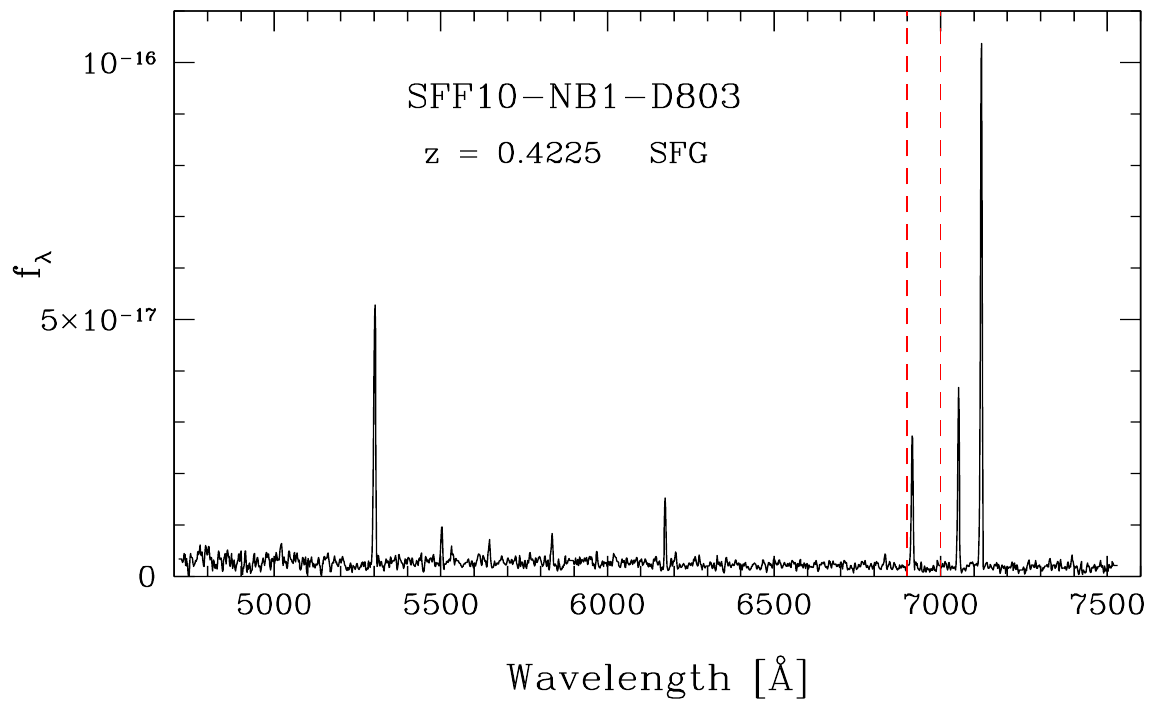} \includegraphics[width=0.49\textwidth,keepaspectratio]{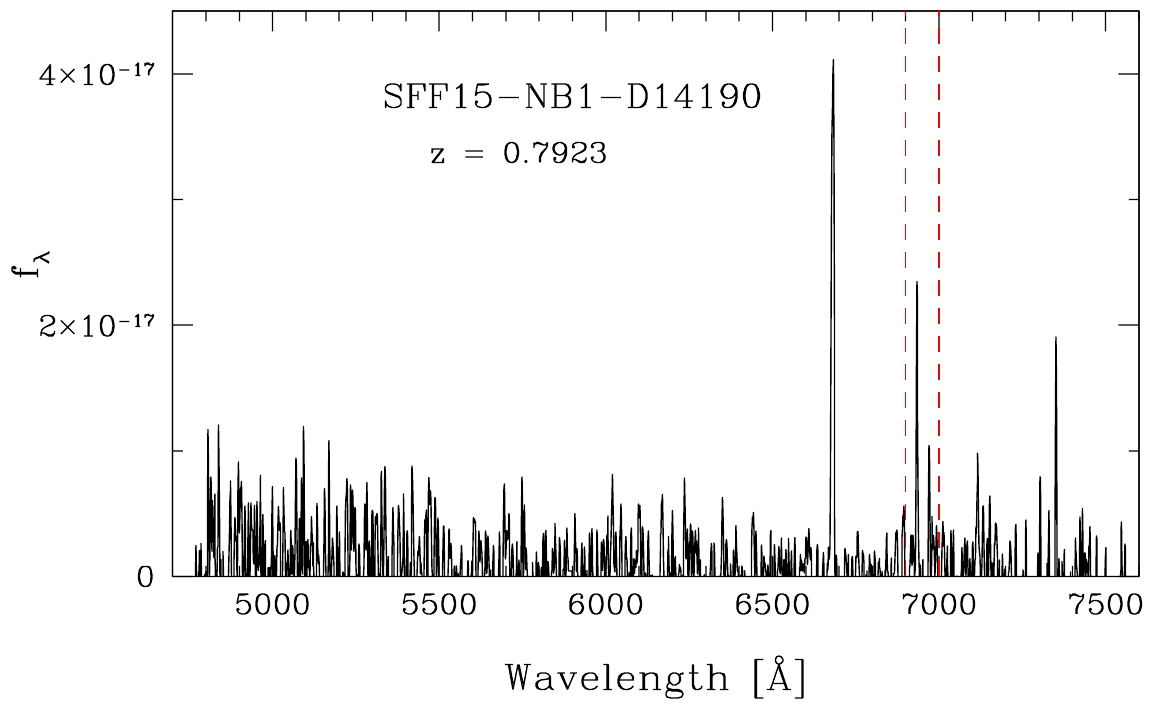}
    \put(-455,130){\large (c)}
    \put(-288,115){\scriptsize [O III]}
    \put(-322,55){\scriptsize H$\beta$}
    \put(-367,52){\scriptsize H$\gamma$}
    \put(-391,44){\scriptsize H$\delta$}
    \put(-434,95){\scriptsize [O II]}
    \put(-205,130){\large (d)}
    \put(-45,93){\scriptsize [Ne III]}
    \put(-92,110){\scriptsize [O II]}

    \caption{Spectra of four galaxies detected by a non-primary emission line (i.e., not their \HA, \OIII, or \OII\ lines). The red-dashed vertical lines denote the wavelength range covered by the narrow-band filter in which the source was detected.   Key emission lines are labelled.  The flux scales on the y-axes are in units of erg s$^{-1}$ cm$^{-2}$ \AA$^{-1}$.
    (a): An object detected by the \SII\ doublet in the NB1 filter.   This object is not from the pilot study.
    (b): A star-forming galaxy detected via \OIII$\lambda$4959 in the NB2 filter.
    (c): An \HB\ detection discovered in NB1.
    (d): An object detected by its \NeIII$\lambda$3869 line in the NB1 filter.
    \label{fig:additional_lines}}
    \end{figure*}
    
       % Additional lines
    In addition to detecting galaxies with the three primary lines that the SFACT survey was expected to detect, objects are also detected via a variety of other emission lines. Though these objects are not the norm for the SFACT survey ($\sim$5\% of non-QSO detections), it is worth highlighting a few examples to demonstrate the types of objects detected with these lines. Figure \ref{fig:additional_lines} presents spectra of some SFACT objects detected by a line other than \HA, \OIII, or \OII. 
    
    Panel (a) of Figure \ref{fig:additional_lines} shows an object detected due to its \SII\ emission. Since there are no \SII\ detections in the pilot-study fields we have included this object from the SFF08 field to illustrate this type of detection. This particular object is an \HII\ region in a metal rich, luminous star-forming spiral galaxy.  Note that \SII\ detections are only possible in the NB1 and NB3 filters, since an \SII\ detection in the NB2 filter would require the object to possess a large blueshift.  Panel (b) of the figure shows an object detected by its \OIII$\lambda$4959 line. There are 16 objects in the pilot study fields that are detected by this line.  As is evident in the figure, the 5007 line has been redshifted out of the filter's wavelength range. Since the 4959 line is significantly weaker than the 5007 line, it makes sense that this is somewhat of an unusual occurrence in the survey fields.  Most of the 4959 detections are strong ELGs with high EW lines, such as SSF15-NB2-A6234 shown here.  Panel (c) shows one of the three \HB\ detections in the pilot-study fields, this one detected in the NB1 filter. It has an absolute magnitude of M$_g = -19.8$ and is likely a Green Pea-like star-forming galaxy. Finally, panel (d) presents the spectrum of the only \NeIII$\lambda$3869 detection in the pilot-study fields. It has a redshift of z~=~0.79 and an absolute magnitude of M$_g = -20.4$.
   
  The fact that SFACT can detect objects via these additional emission lines speaks to the sensitivity of the survey method.  However, a secondary reason for presenting these spectra of ELGs detected via non-standard lines is to inject a note of caution regarding NB surveys.   Most such surveys with depths comparable to or greater than that of SFACT will typically not possess follow-up spectra for most of their ELG candidates.  It is clear that a sensitive NB survey can easily detect galaxies via their \SII\ lines that might well be mistaken for \HA-detections.  Alternatively, such surveys will likely detect numerous galaxies via their \HB\  or \OIII$\lambda$4959 lines, and mistake them for \OIII$\lambda$5007 detections.   While the results presented here suggest that the level of such contamination should be small, it is likely to increase with the depth of the NB survey.

\subsubsection{Spectra of SFACT QSOs} \label{sec:additionalspec}

    \begin{figure*}[t]
    \centering
    \includegraphics[width=0.49\textwidth,keepaspectratio]{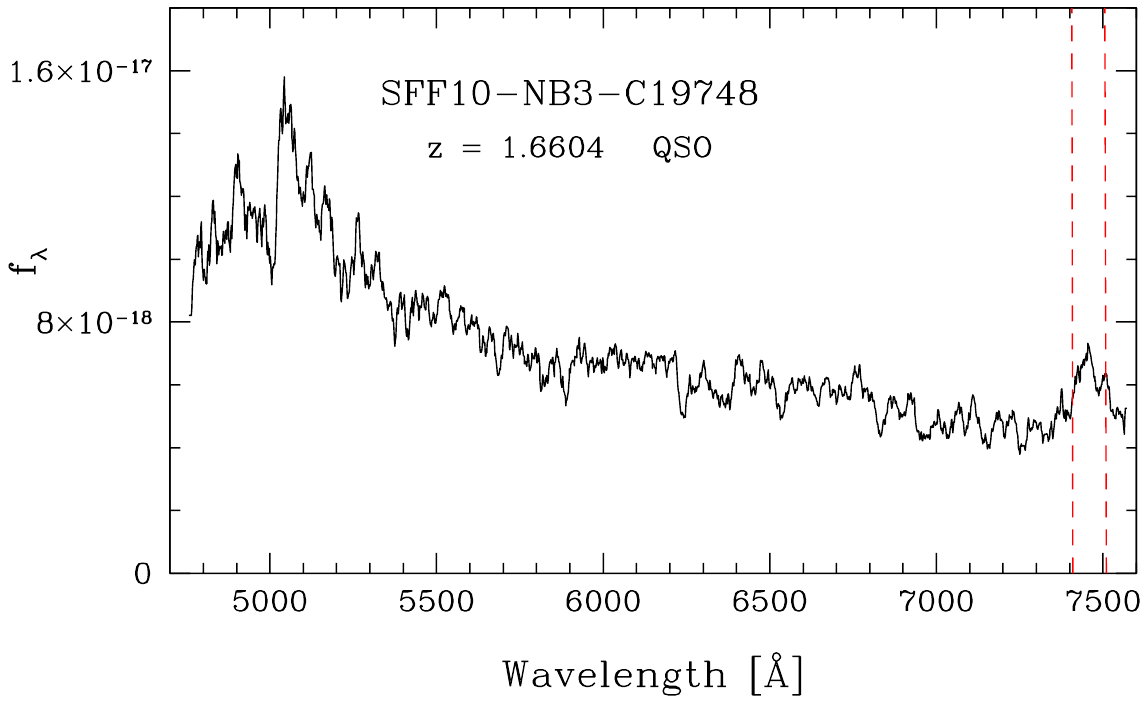} \includegraphics[width=0.49\textwidth,keepaspectratio]{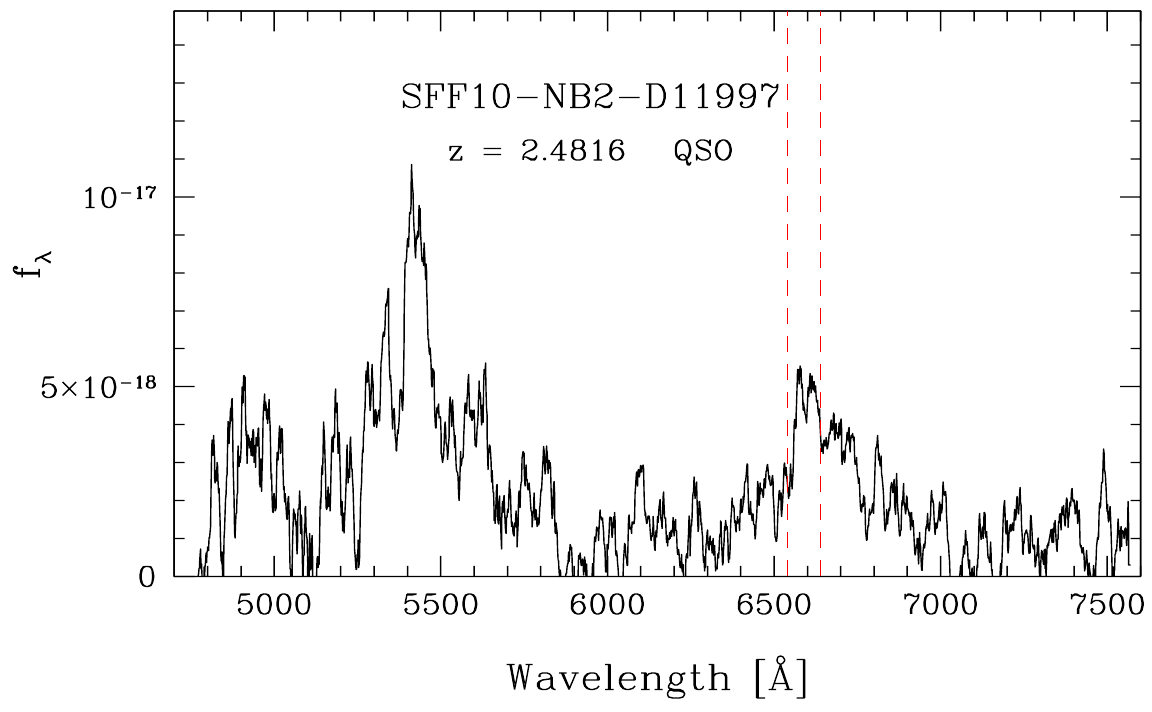} 
    \put(-460,130){\large (a)}
    \put(-295,80){\scriptsize Mg II}
    \put(-450,75){\scriptsize C III]}
    \put(-205,130){\large (b)}
    \put(-68,75){\scriptsize C III]}
    \put(-187,110){\scriptsize C IV}
     
    \includegraphics[width=0.49\textwidth,keepaspectratio]{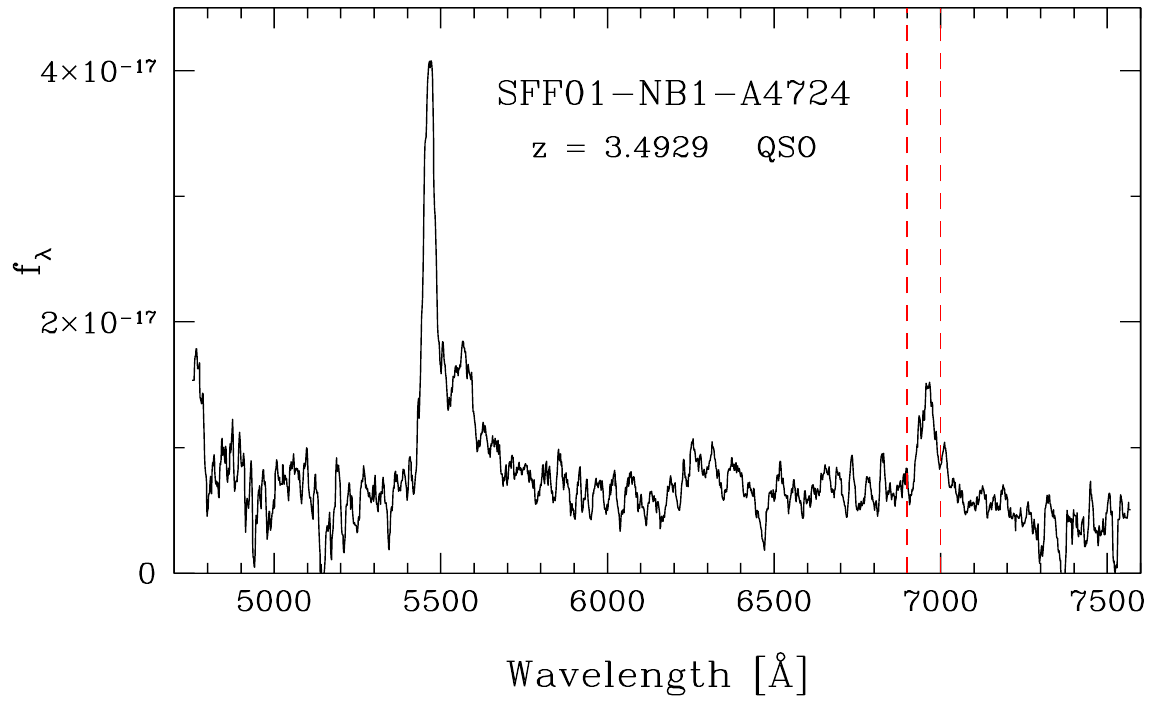} \includegraphics[width=0.49\textwidth,keepaspectratio]{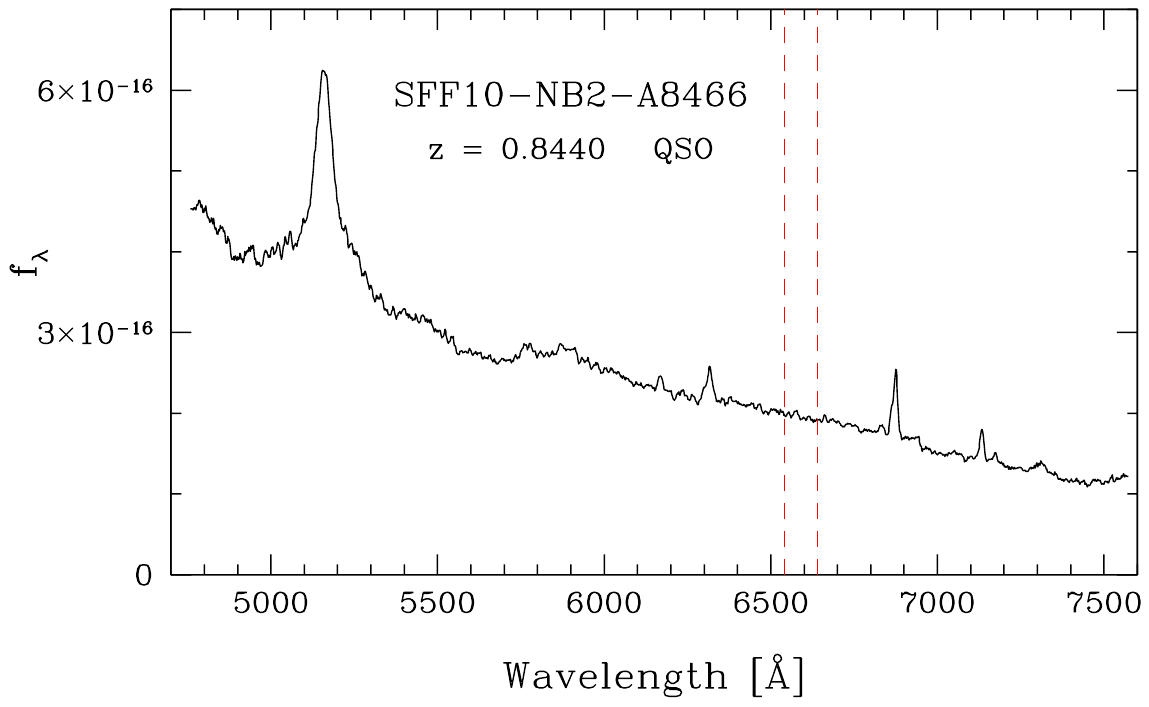} 
    \put(-460,130){\large (c)}
    \put(-297,70){\scriptsize C IV}
    \put(-432,120){\scriptsize Ly$\alpha$}
    \put(-205,130){\large (d)}
    \put(-49,65){\scriptsize [Ne III]}
    \put(-65,80){\scriptsize [O II]}
    \put(-112,80){\scriptsize [Ne V]}
    \put(-192,80){\scriptsize Mg II}
      
    \caption{Spectra of four QSOs detected in the pilot study. The red-dashed vertical lines denote the wavelength range covered by the narrow-band filter in which the source was detected.   Key emission lines are labelled.  The flux scales on the y-axes are in units of erg s$^{-1}$ cm$^{-2}$ \AA$^{-1}$.
    (a): A QSO detected via Mg~{\footnotesize II} $\lambda 2798$ in NB3. C~{\footnotesize III]}$\lambda 1908$ is also visible.
    (b): A C~{\footnotesize III]}$\lambda 1908$, NB2 detected QSO, with C~{\footnotesize IV} $\lambda$1549 also visible.
    (c): A QSO detected by C~{\footnotesize IV} $\lambda$1549 in the NB1 filter. Ly$\alpha$ is also visible.         
    (d): A color-selected quasar detected in NB2 without any emission line present in the filter. See text for explanation on how this object was detected. 
\label{fig:additional_QSOs}}
    \end{figure*}
    
     %QSOs
    While most of the emission-line detections in SFACT are of star-forming galaxies, there are also thirteen QSOs found in the pilot-study fields to date.  From the outset of the project, we anticipated finding modest numbers of line-selected QSOs in our NB images.  The expectation was that we would detect QSOs via one of the common UV emission lines, such as \ion{Mg}{2} $\lambda$2798, \ion{C}{3}] $\lambda$1908, \ion{C}{4} $\lambda$1549, or Ly$\alpha$ $\lambda$1215.  In addition to the quasar presented in Figure \ref{fig:Quasar}, we present four additional example spectra of some of these objects in Figure \ref{fig:additional_QSOs}, illustrating detections in each of the first three lines listed above.   There are no Ly$\alpha$-detected QSOs in the pilot-study fields, but several have been cataloged in other SFACT fields and will be presented in subsequent catalog papers.
    
    Panel (a) of Figure \ref{fig:additional_QSOs} shows a quasar with a Mg~{\footnotesize II} $\lambda 2798$ emission line that was redshifted into the NB3 filter. It has a redshift of z~$= 1.66$ and an absolute magnitude of M$_g = -24.4$.  Panel (b) shows a C~{\footnotesize III]} $\lambda 1908$-detected QSO discovered with the NB2 filter. It has an absolute magnitude of M$_g = -23.2$ and a redshift of z~$= 2.48$.  It is one of the faintest of the SFACT-detected QSOs, with an r magnitude of 22.85; this helps to explain the poor quality of its spectrum.   Panel (c) of the figure shows a quasar that was detected by its C~{\footnotesize IV} $\lambda$1549 in the NB1 filter. Its redshift is z~$= 3.49$ and its absolute magnitude is M$_g = -26.2$. This object is the second highest redshifted object in the pilot study, surpassed only by another C~{\footnotesize IV}$\lambda$1549-detected quasar at z~$= 3.51$.  In all of the spectra shown in Figure 7, at least one additional UV emission line is present, making the line identifications and redshift measurements secure.

    Panel (d) of the figure shows a quasar that strongly departs from the properties of the others.   While is was ``detected" in the NB2 filter,  our spectrum reveals that no line is present in the wavelength range covered by that filter.  This happens occasionally when QSOs with strong, steep blue continua get selected by the automated software even though they have no line in the narrow-band filter. We call these objects color-selected QSOs and they mostly occur in the NB2 filter.  There are only two such objects detected among the pilot-study fields, but many other examples exist in the other SFACT fields.  In the example shown here, the measured redshift is z = 0.844, the r magnitude is 19.50 (the brightest of the 13 SFACT QSOs in the current sample), and it has an absolute magnitude of M$_g = -23.8$. 
    
    The reason these objects are detected in the survey has to do with our selection methodology. For the NB2 filter, the \textit{r} and \textit{i} filters are summed together to form the broadband continuum image. These filters are roughly equivalent to the SDSS filters and span the $5600-8200$~\AA\ range. The NB2 filter is located at 6590~\AA, closer to the blue end of this wavelength range. For a QSO like the one in Figure \ref{fig:additional_QSOs}(d), the strongly blue-sloping continuum is much higher at the location of the NB2 filter than it is at middle of the continuum filter (around 7000~\AA).  Therefore, for an almost purely continuum source, more flux is measured at NB2 than from the scaled continuum image at 7000~\AA. When the NB and continuum fluxes are compared, there is excess flux in the NB filter that is detected by the software. For more information on the imaging data and the specifics of the survey's selection methods see \citetalias{Paper2}.
    
    For the purposes of the SFACT survey, we will catalog the color-selected QSOs that are selected by our software.  Strictly speaking, the color-selected QSOs are NOT emission-line objects, but since they represent {\it bona fide} NB detections, it seems appropriate to retain them in our catalogs.  They will be treated differently, however, from the line-selected QSOs (which are substantially more common), and not used in any statistical studies of the SFACT QSOs.

%************************************************************************************************************************

\begin{deluxetable*}{cccccc}
%    \tabletypesize{\scriptsize}
   \tablecaption{Summary of Detected Emission Lines in Pilot-Study Fields\label{tab:lines}}
    \tablehead{
    \\
     Line & \# of NB1 & \# of NB2 & \# of NB3 & Total, All & Object Description  \\
    & Detections & Detections & Detections & Filters &   \\  %$\frac{\text{erg}}{\text{s } \text{cm}^2}$
    (1) & (2) & (3) & (4) & (5) & (6) 
    }
    \startdata
\ \ion{C}{4} $\lambda$1459    &  3  &  0  &   0 &  3  & QSOs - z range 3.22 to 3.85 \\
\ \ion{C}{3}] $\lambda$1908   &  0  &  2  &   0 &  2  & QSOs - z range 2.43 to 2.94 \\
\ \ion{Mg}{2} $\lambda$2798 &  1  &  3  &   2 &  6  & QSOs - z range 1.34 to 1.69 \\
\\
\ [\ion{O}{2}]  $\lambda$3727 & 40 & 26  &  29 & 95  & ELGs - z range 0.75 to 1.02 \\
\ [\ion{Ne}{3}]  $\lambda$3869 &  1 &  0  &   0 &  1  & ELGs - z range 0.69 to 0.95 \\
\\
\ H$\beta$  $\lambda$4861 &  2 &  0  &   1 &  3  & ELGs - z range 0.34 to 0.55 \\
\ [\ion{O}{3}]  $\lambda$4959 &  3 &  9  &   4 & 16  & ELGs - z range 0.32 to 0.52 \\
\ [\ion{O}{3}]  $\lambda$5007 & 47 & 62  &  53 & 162  & ELGs - z range 0.30 to 0.50 \\
\\
\ H$\alpha$ $\lambda$6563 -- All & 53 &  2  &  70 & 125  & ELGs + \ion{H}{2} regions - z range 0.00 to 0.15 \\
\ H$\alpha$ detected \ion{H}{2} regions & 12 &  1  &  4 & 17  &  \ion{H}{2} regions - z range 0.00 to 0.15 \\
\ Host Galaxies w/ \ion{H}{2} regions & 21 &  1  &  14 & 36  &  Host Galaxies - z range 0.00 to 0.15 \\
 \enddata
%    \tablecomments{}
    \end{deluxetable*}

\subsection{Properties of the SFACT Galaxies} \label{sec:properties}

A preliminary evaluation of the properties of the SFACT galaxies detected in the pilot-study fields is presented in SFACT1.   Here we provide an updated view of some of these properties, based on the results of the spectroscopic analysis presented in the current paper.  In particular, we present updated redshift histograms in Section~\ref{sec:redshiftdist} and substantially enhanced emission-line diagnostic diagrams that make use of our re-examination measurements in Section~\ref{sec:emissionlinediagnosticdiagrams}.
    
Table~\ref{tab:lines} provides a summary of the specific emission lines used to detect each of the SFACT galaxies in the current study.  Only spectroscopically-verified objects are included.  The first column of the table lists the emission lines used to detect our sources, while the numbers of detections for each of these lines is indicated for each of the NB filters individually (columns 2-4) and as a total number (column 5).  The last column specifies the type of objects being detected and the redshift ranges relevant for each line.

As mentioned in the previous section, SFACT is capable of detecting objects via multiple emission lines.   Objects are detected via a total of nine different lines in these three survey fields.  Our broader sample also includes additional lines that do not happen to be represented in the current sample (e.g., [\ion{S}{2}] $\lambda\lambda$6717,6731, H$\gamma$ $\lambda$4340, and Ly$\alpha$ $\lambda$1215).  

Not surprisingly, the frequency with which SFACT detects a given line varies substantially.  Among the z $\lesssim$ 1.0 ELGs there is an obvious tendency for the survey to detect objects via the stronger lines: H$\alpha$ (n = 125), [\ion{O}{3}] $\lambda$5007 (n=162), and [\ion{O}{2}] $\lambda$3727 (n=95).  This is no surprise, and is consistent with our expectations.  These three lines are the primary lines detected by SFACT.  The other optical lines like [\ion{Ne}{3}]  $\lambda$3869, H$\beta$  $\lambda$4861, and [\ion{O}{3}]  $\lambda$4959 occur much less frequently.  It is worth noting that for some redshifts, {\it both} of the [\ion{O}{3}] lines are located within the NB filter, which will enhance the probability for detection.  These cases are all classified as [\ion{O}{3}] $\lambda$5007 detections, since this line is the stronger of the two.   Future survey papers employing larger samples will explore the impact of this selection-function enhancement.

In Table~\ref{tab:lines} we break out H$\alpha$ detections that are flagged as \ion{H}{2} regions and as host galaxies of \ion{H}{2} regions.  SFACT1 details the reasons for distinguishing between these two categories of objects.  In short, any object that is detected in SFACT as containing an \ion{H}{2} region outside of the nucleus is labeled as a host galaxy.   In some, but not all, cases one or more bright \ion{H}{2} regions may also be cataloged (appropriately linked to their host galaxy).  This is to allow abundance measurements, particularly in cases where there is little or no emission associated with the nucleus.  In the pilot-study fields there are 36 galaxies identified as hosting \ion{H}{2} regions, all detected via the H$\alpha$ line, and 17 additional \ion{H}{2} regions.  The remaining H$\alpha$ detections are stand-alone ELGs, which are predominantly dwarf star-forming galaxies (see SFACT1).

A total of eleven line-selected QSOs are also included in Table~\ref{tab:lines}.  Here there are less extreme variations between the numbers of objects detected by the three UV lines.  There is an indication that the \ion{Mg}{2} $\lambda$2798 line is favored over the others, which is perhaps no surprise since objects detected via this line are located at smaller distances and hence would appear brighter for a fixed QSO line luminosity.   The small numbers of QSOs present in the current sample precludes a detailed assessment of the QSO population at this time.

As the imaging and spectroscopic observations of additional SFACT survey fields are completed, we will have access to thousands of ELGs with data similar in nature to those presented in the pilot-study papers.   Future SFACT papers will explore the properties of the survey constituents, as well as their completeness limits and selection function, in substantial detail.

\subsubsection{Redshift Distributions} \label{sec:redshiftdist}

\begin{figure}[t!]
    \centering
    \includegraphics[width=\columnwidth,keepaspectratio]{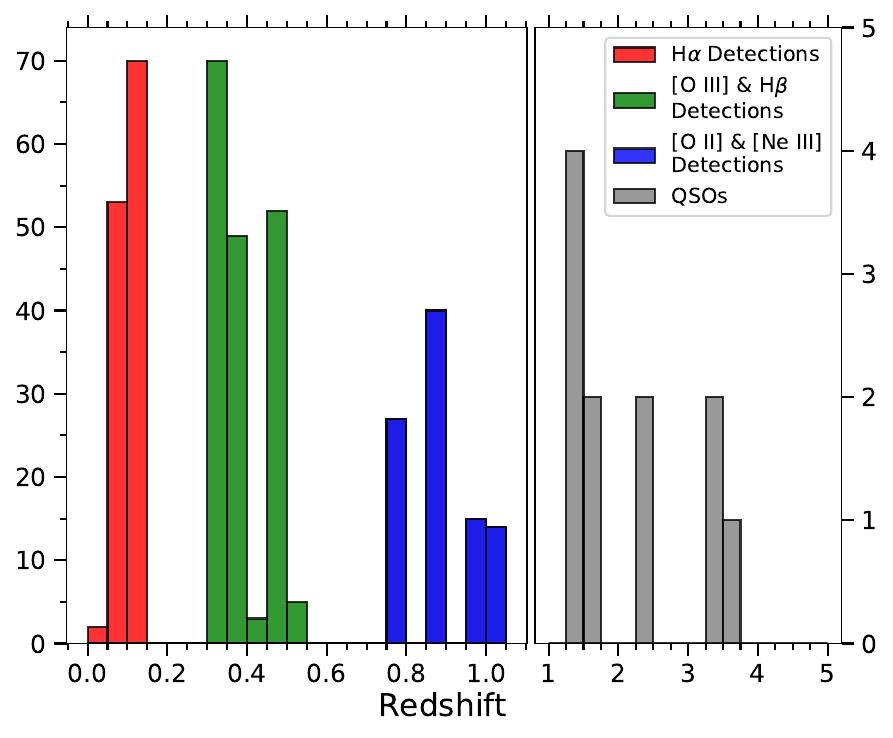}
    \caption{Histogram of SFACT ELGs binned by redshift. SFACT samples the universe in discrete redshift windows with each line-filter combination probing a different redshift range. The bin width changes between the left and right sections of the plot. Bins in the left panel have a widths of 0.05 and bins on the right have widths of 0.25. \label{fig:zhist}}
    \end{figure}

    %Redshift distribution explanation
    A histogram of the redshifts of all the objects in the pilot-study fields are shown in Figure \ref{fig:zhist}.  The figure is broken into two subplots. The left subplot shows the SFACT ELGs that are detected below redshifts of 1.05 and the right shows the SFACT line-detected QSOs from redshifts of 1.25 and beyond. The binning of the two subplots changes, with the left plot having bins of width 0.05 and the right having bins of width 0.25. 
    
    %First group
    There are four groups of objects in Figure \ref{fig:zhist}, each coded by a different color.   Each group represents a portion of the pilot-study sample detected by one or more specific emission lines.   The SFACT survey discovers galaxies within discrete redshift windows, where each window is probed through a different narrow-band filter and each filter can detect different lines at different redshifts.   Each ``peak" in the figure represents a different redshift window of the SFACT survey where one of the primarily emission lines falls within one of the NB filters. 
    
    The first group are the lowest-redshift objects: the \HA\ detected galaxies. Unlike the redshift histogram in SFACT1, Figure~\ref{fig:zhist} includes the \ion{H}{2} regions as well as the Host galaxies and ELGs.  The first redshift bin in this group contains the objects discovered by the NB2, 6590~\AA\ filter which spans a redshift range of $-0.002-0.011$, the second bin includes the objects discovered by the NB1, 6950~\AA\ filter which spans redshifts $0.052-0.066$, and the third bin represents galaxies discovered by the NB3, 7450~\AA\ filter across redshifts $0.129-0.144$.   All \HA-detected histogram bins are shown in red.  The population of the first bin in the \HA-detected objects is small, as expected due to the small volume of space searched in the NB2 filter. The other two narrow-band filters yielded higher quantities of \HA\ detections.

    %Image of Redshift distribution

    %second group
        The second grouping of bins in Figure \ref{fig:zhist} show primarily \OIII$\lambda 5007$-detected galaxies, shown in green. %As opposed to the \HA-detected ELGs, the number of objects from each filter are more consistent. 
For the \OIII$\lambda 5007$ line, the NB2 filter probes a redshift range of $0.308-0.325$ which places these sources within the first green bin. The second bin lines up with the NB1 redshift range of $0.378-0.397$ and the fourth bin includes the NB3 redshift range of $0.480-0.500$. In total, the \OIII$\lambda 5007$-selected sample contains 162 objects and represents the largest sample in the SFACT pilot study.  This section of the redshift histogram also contains \OIII$\lambda 4959$ and \HB-detected ELGs. The \OIII$\lambda 4959$-detected objects have their $\lambda 5007$ line redshifted out of the filter so that only the $\lambda 4959$ line is captured in the filter's wavelength range. Since the 4959 line is significantly weaker than the 5007 line, we expect to detect fewer galaxies this way. This holds true as we only have 16 ELGs that are detected via the \OIII$\lambda 4959$ line. %The various objects detected with \OIII$\lambda 4959$ fall in the second, third, and fifth bins in this portion of the histogram. Additionally, 
There are also three \HB-detected ELGs in this portion of the histogram.       
    %third group
    The final group of objects in the left panel of Figure \ref{fig:zhist} are primarily detected by their \OII$\lambda 3727$ doublet. These sources are shown in blue.   The \OII\ line is detected in the redshift range of 0.757$-$0.780 in the NB2 filter, in the range 0.852$-$0.877 in NB1, and 0.988$-$1.015 in NB3.  The latter group of detections is split between two histogram bins in Figure \ref{fig:zhist}.
%   The redshift range for the \OII-detected objects in the pilot study is $0.756 <$~z~$< 1.013$ so, 
At these distances all of the detected objects are unresolved in our imaging data and appear as dots. They are also some of the faintest objects in the survey. A single source is detected due to its \NeIII$\lambda 3869$ line in the NB1 filter at z~=~0.79. 
        
    %QSOs
    %remake QSO plot with updated ZS
    Finally, there are a total of 11 line-selected QSOs with spectroscopic follow up in the SFACT pilot-study fields. These objects are shown in the right panel of Figure \ref{fig:zhist}. Six were detected by Mg~{\footnotesize II}$\lambda 2798$ with redshifts between 1.34 and 1.67, two were detected by C~{\footnotesize III}]$\lambda 1908$ with redshifts 2.47 and 2.48, and three were detected by the C~{\footnotesize IV}$\lambda 1549$ line with redshifts between 3.48 and 3.51. This is a small number of detections compared to the number of ELGs discovered by SFACT.   However, given the relatively small volumes of the three pilot-study fields, combined with the relative rarity of QSOs, the small number of QSOs is not unexpected.   For the completed SFACT survey ($\sim$ 60 total  fields), we expect to detect about 200$-$300 QSOs located in the survey redshift windows.

\subsubsection{Emission-Line Diagnostic Diagrams} \label{sec:emissionlinediagnosticdiagrams}
    
    Emission-line diagnostic diagrams are used to separate galaxies based on their ionization sources (star-forming or AGN) and are often used to reveal some of the physical conditions present in these galaxies \citep{1981PASP...93....5B, 1987ApJS...63..295V}. The most famous of these emission-line diagnostic diagrams is the Baldwin, Philips, Terlevich (BPT) diagram which uses the \OIII/\HB\ ratio vs. the \NII/\HA\ ratio to separate star forming galaxies from their AGN counterparts. In addition to the BPT diagram, \citet{1981PASP...93....5B} also present an \OIII/\HB\ vs. \OII/\OIII\ diagram. The redshift ranges and spectral coverage of the survey means that the SFACT galaxies only have certain emission lines present in their spectra and different lines will fall outside the wavelength range of the follow-up spectra depending on which  emission line was detected in the SFACT narrow-band filters. As a result, we will display the SFACT pilot-study sample on both the classic BPT diagram as well as the \OIII/\HB\ vs. \OII/\OIII\ diagram.  
    
    Before these objects can be added to any emission-line diagnostic diagram, their emission-line ratios need to be corrected for underlying Balmer absorption and reddening when possible. In the case of the underlying absorption,  we adopt a statistical correction of 2~\AA\ of underlying absorption for each Balmer line, which is consistent with  the correction in \citet{1993ApJ...411..655S} and \citet{2022ApJ...925..131H}. The corrected Balmer lines are then used to determine the reddening correction, c$_{\text{\HB}}$, following the standard procedure \citep{2006agna.book.....O} and use \HA\ and \HB\ for \HA-detected sources and \HB\ and \HG\ for \OIII-detected sources. In cases where Balmer lines are not  detected or the computed value of c$_{\text{\HB}}$ comes out negative, c$_{\text{\HB}}$ is set to 0.  The derived values for c$_{\text{\HB}}$ range between 0.00 and 1.77 for the current sample, with a mean value of 0.25.  The corrected emission-line ratios shown in the plots in this section are the same ratios presented in the last columns of Tables \ref{tab:specdata_sfactf01} through \ref{tab:specdata_sfactf15}.
    
    For the \HA-detected sample, our spectral wavelength coverage does not extend to the \OII\ doublet and \SII$\lambda \lambda 6731,6716$ falls outside the survey's redshift range for NB3 detected objects. Therefore, the BPT diagram is the best option for plotting these objects on an emission-line diagnostic diagram. However, only 56 of the 125 \HA-detected sources have the four required lines automatically detected by {\tt WRALF}. Many of these objects have \NII\ or \HB\ lines that fall below the signal-to-noise ratio required by the software to automatically detect them and must be re-examined. As described in Section \ref{sec:wralf}, we look at the \HA-detected sources that are missing these lines and determine if there is a line present at the expected location for \NII\ or \HB. These emission lines are re-examined and are labeled as Category 1 or Category 2 measurements, depending on the SNR of the line.  Based on these additional measurements, we have added 52 objects to the BPT diagram for a total of 108 of the pilot study's 125 \HA-detected objects.   Note that objects that possess lower quality Category 2 measurements for both the \NII\ and \HB\ lines have been excluded from the plot. 
    
    %BPT
    \begin{figure}[t]
    \centering
    \includegraphics[width=\columnwidth,keepaspectratio]{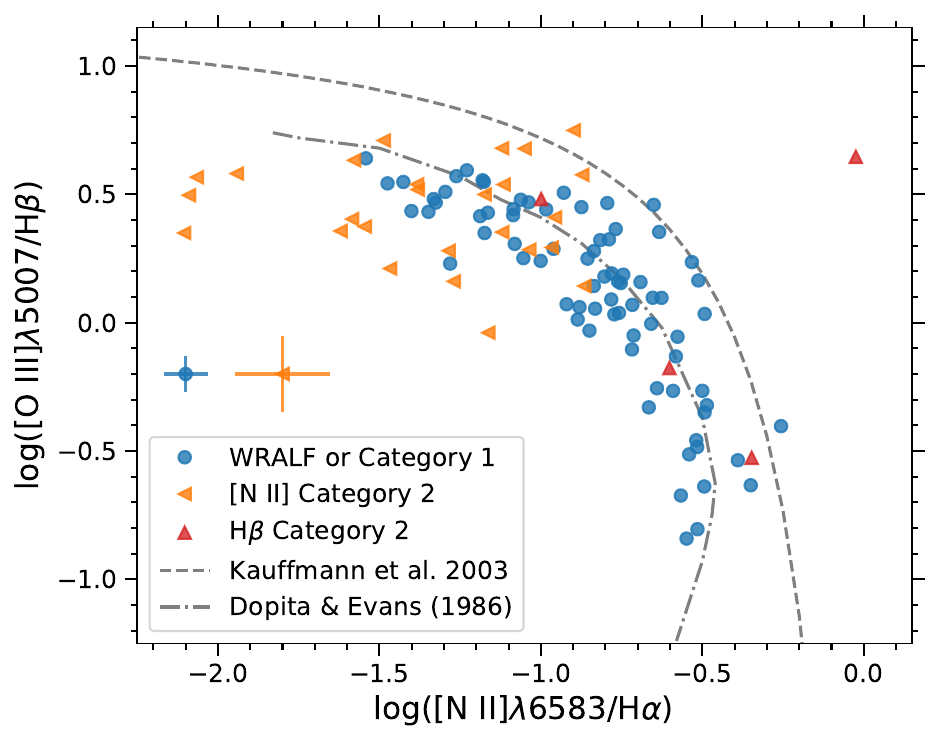}
    \caption{We present the BPT diagram for the \HA-detected pilot-study sample. Objects measured by {\tt WRALF} or re-examined and labeled as a Category 1 measurement are displayed as blue circles. Objects with the \NII\ or \HB\ line re-examined and labeled as a Category 2 measurement are shown as left pointing, orange triangles or upward pointing, red triangles. Objects that are a Category 2 measurement for both the \NII\ and \HB\ re-examination have been excluded.  Characteristic error bars for the two groups of objects (WRALF+Category 1 or Category 2) are shown.  The dashed-dotted line is from \citet{1986ApJ...307..431D} and is derived from stellar photo-ionization models. The dashed line is from \citet{2003MNRAS.346.1055K} and is an empirical deliminator between the star-forming galaxies and AGN. \label{fig:BPT}}
    \end{figure}
    
    The \HA-detected galaxies are plotted on Figure \ref{fig:BPT}. Here blue circles are objects that were measured by {\tt WRALF} or are a Category 1 measurement of either \NII\ or \HB\ (or both). Category 2 objects with the \NII\ or \HB\ line re-examined are shown as left pointing, orange triangles or upward pointing, red triangles respectively. In the cases where the re-examined line measurement is an upper limit, the direction the triangle points indicates the direction these objects could move on the BPT diagram if we were able to more precisely measure the emission line with a higher signal-to-noise observation of these objects. In total, there are 26 objects with Category 2 \NII\ emission lines and 4 objects with Category 2 \HB\ emission lines.
    
    The empirical \citet{2003MNRAS.346.1055K} line is indicated by the dashed line and separates the SF and AGN components of the diagram. The dashed-dotted curve that goes through the star-forming section of the graph is from \citet{1986ApJ...307..431D} and represents the high-excitation stellar photo-ionization models from their work. Most of the sources discovered by SFACT are star-forming galaxies and fall under the \citet{2003MNRAS.346.1055K} line. Additionally, many of the objects on the left side of the diagram have Category 2 \NII\ emission line measurements. This is because \NII\ is typically very weak in metal-poor galaxies. The objects on the far left of the diagram have weakly determined \NII\ lines and, in some cases, the re-examination measurement is essentially an upper limit of this line's flux. There is one potential Seyfert 2 galaxy present in the upper right portion of Figure \ref{fig:BPT}, although this is a system with Category 2 line measurements and hence its line ratios are highly uncertain. There is also a potential LINER that falls close to the \citet{2003MNRAS.346.1055K} line in the bottom right.
    
    %o3hb v o2o3
    \begin{figure}[t]
    \centering
    \includegraphics[width=\columnwidth,keepaspectratio]{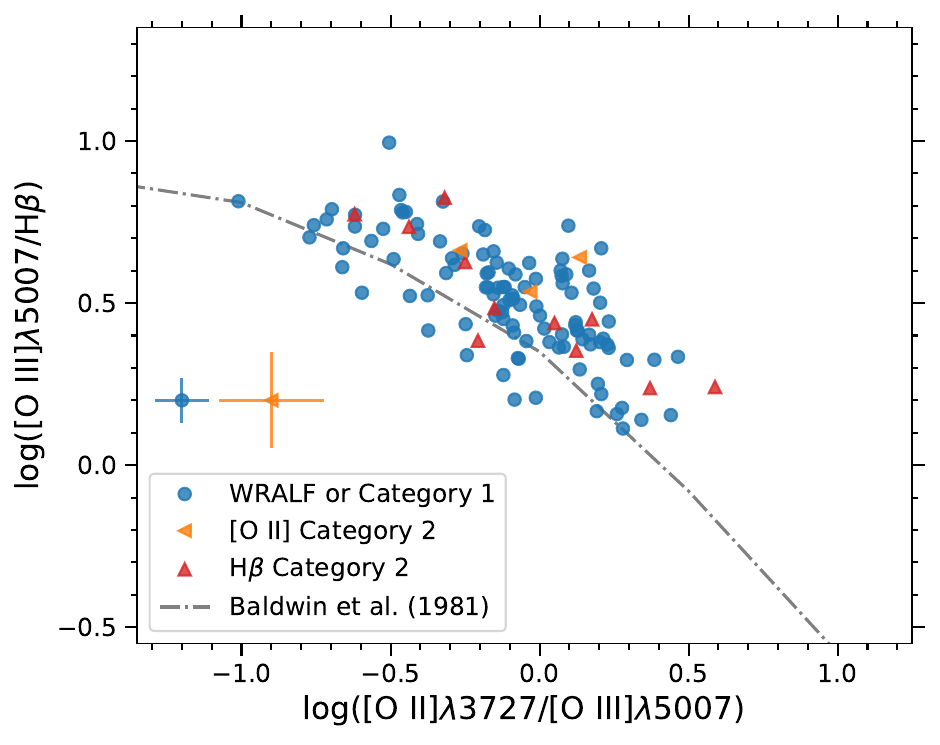}
    \caption{We present the \OIII-detected sample plotted on a $\log$(\OIII/\HB) vs. $\log$(\OII/\OIII) emission-line diagnostic diagram. The symbols are as in Figure \ref{fig:BPT} except this time orange left pointing triangles are \OII\ Category 2 re-examination measurements.  Characteristic error bars for the two groups of objects (WRALF+Category 1 or Category 2) are shown.   The dashed-dotted line is from \citet{1981PASP...93....5B} Figure 1 and fits emission-line ratios of approximately solar metallicity \HII\ regions. \label{fig:diag2}}
    \end{figure}
    
    For the \OIII-detected sample, none of the lines around \HA\ are present in our spectral wavelength coverage. However, the \OII\ line is within the survey's wavelength range for these objects. Thus, the $\log$(\OIII/\HB) vs. $\log$(\OII/\OIII) diagram from \citet{1981PASP...93....5B} was chosen to display the \OIII-detected galaxies and is shown in Figure \ref{fig:diag2}. The symbols in this figure are the same as Figure \ref{fig:BPT} except this time orange left pointing triangles are \OII\ Category 2 re-examinations. Only 95 of the 179 objects have all the lines automatically detected by the software with 27 added by the re-examining process for a total of 122 objects on this plot. Of those 27, 13 are Category 1 re-examinations. For the 14 remaining Category 2 measurements, 3 have \OII\ measurements and 11 are \HB\ measurements that are flagged as Category 2 re-examinations. 
    
    The dashed line in this figure shows the trend line from Figure 1 in \citet{1981PASP...93....5B} and fits emission-line ratios of approximately solar metallicity \HII\ regions and planetary nebulae. The fact that many of the galaxies are above this line implies they have lower abundances. This, taken with the fact that they also have lower luminosity (see \citetalias{Paper1}) implies that they would occupy the upper left portion of the star-forming sequence in Figure \ref{fig:BPT}.  A single confirmed Seyfert 2 galaxy is plotted in this figure as well. This object, whose spectrum is shown in Figure \ref{fig:OIII}a, falls above the trend line at log(\OIII/\HB) of about 1.0. It is possible that some of the other galaxies plotted here are also Seyfert 2 AGN, but without their \NII/\HA\ ratios they can't be easily separated from their star-forming counterparts.
    
    Finally, the \OII-detected sample has an extremely limited number of lines in the survey's spectral coverage. For this reason, they have not been plotted on any emission-line diagnostic diagram. In the future, we plan to reobserve the \OII-detected galaxies with spectral coverage that probes deeper into the red part of the spectrum, allowing additional lines to be measured for these objects so they can be placed similarly on these graphs.

\section{Summary and Conclusions} \label{sec:summary}

The current paper, the third of a series, presents the spectroscopic data from the SFACT Survey pilot-study fields SFF01, SFF10, and SFF15 and describes how the spectroscopic portion of the survey is being carried out.   Previous papers in this series present an overview of the overall survey \citepalias{Paper1} and describe the imaging portion of SFACT \citepalias{Paper2}.  The primary goals of the spectroscopic component of the SFACT program include the verification of objects cataloged in our narrowband images, including providing feedback on the candidate selection process, and the measurement of key properties such as activity class (e.g., star forming vs. AGN), accurate redshifts, and physical  characteristics such as internal absorption and metal abundances.

Using the WIYN 3.5m telescope, as well as the multi-fiber positioner Hydra and the Bench Spectrograph, the survey is able to carry out follow-up spectroscopy on a large number of potential ELG candidates. Our chosen wavelength range of $4760-7580$~\AA\ allows spectra to be simultaneously obtained for all objects detected in the three narrow-band filters of the SFACT survey.   Each multi-spectral image is run through various processing steps after which emission lines within the spectra are identified and measured through the use of software in a semi-automatic fashion.  Any lines missed by the software are re-examined and added into the survey data tables to extract as much useful information from the spectra as possible.

Example objects are shown for each of the primary emission lines used to detect objects in the survey (\HA, \OIII$\lambda 5007$, and \OII$\lambda 3727$) in each of the narrow-band filters.   Spectra of additional objects detected by non-primary emission lines are also shown, as are spectra of several quasars detected in the pilot study. These objects demonstrate the wide range of objects in the SFACT catalog as well as the power and versatility of the SFACT survey to detect emission lines in even the faintest of sources. The redshift distribution of the pilot study is also examined, showing the range of distances at which these objects are detected. Finally, the sample is placed on two different emission-line diagnostic diagrams to distinguish the star-forming galaxies from the AGN.

The 533 SFACT sources detected in the three pilot-study fields are tabulated and spectral information for the 453 objects that have follow-up spectra is presented.  415 of these 453 are confirmed emission-line objects, giving the pilot-study a 91.6\% successful detection rate for ELGs. This rate is expected to go up for the later SFACT fields as the survey's methodology has improved greatly over time. 

As of the writing of this paper, spectra have been obtained for many additional SFACT survey fields over the course of several observing runs (November 2017 through March 2022). In total we have followed up on approximately 3800 potential ELGs.  There are currently 35 additional SFACT fields with complete imaging data, most of which have some amount of spectroscopic follow-up.. Furthermore, these fields have the benefit of improvements made in our survey methods based on lessons learned during the analysis of the pilot-study fields. With hundreds of additional SFACT targets awaiting processing and measurement, future papers will have a significantly improved sample size with which to conduct interesting analysis. 

\begin{acknowledgments}

We would like to thank the College of Arts and Sciences at Indiana University for their long term support of the WIYN Observatory. The Department of Astronomy and the Office of the Vice Provost for Research at Indiana University have also helped to support this project with additional funds. 

We would like to thank the anonymous referee who made many substantial and insightful suggestions which improved the final version of this manuscript.

We want to thank the entire staff of the WIYN Observatory, who have made this survey possible. In particular, we acknowledge Daniel Harbeck, Wilson Liu, Susan Ridgeway, and Jayadev Rajagopal. We also thank Ralf Kotulla (U. Wisconsin) for his development and continued support of the ODI image processing software (QuickReduce), and Arvid Gopu and Michael Young (Indiana U.) for their support of the ODI Pipeline, Portal and Archive. Additionally, we wish to thank the WIYN telescope operators without whom we would not have been able to collect our data. 

We would like to acknowledge the efforts made by various students in the Department of Astronomy at Indiana University who have all contributed to the data processing at various points in the project: Anjali Dziarski, Sean Strunk, and John Theising. 

Finally, the authors are honored to be permitted to conduct astronomical research on Iolkam Du'ag (Kitt Peak), a mountain with particular significance to the Tohono O'odham.
\end{acknowledgments}


\vspace{5mm}
\facility{WIYN:3.5m}

\vspace{5mm}
\software{{\tt IRAF}, {\tt WRALF} \citep{Cousins}, and {\tt ALFA} \citep{2016MNRAS.456.3774W}}

\bibliography{mybib}{}
\bibliographystyle{aasjournal}

\end{document}